\documentstyle[epsf,times]{mn2e}

\newcommand{\simgt}{\lower.5ex\hbox{$\; \buildrel > \over \sim \;$}}
\newcommand{\simlt}{\lower.5ex\hbox{$\; \buildrel < \over \sim \;$}}
\newcommand{\bm}[1]{\mbox{{\it \boldmath$#1$}}}
\newcommand{\kaco}[1]{\left\langle{#1}\right\rangle}
\newcommand{\skaco}[1]{\langle{#1}\rangle}

\newcommand{\apj}{ApJ}
\newcommand{\mnras}{MNRAS}
\newcommand{\LCDM}{$\Lambda$CDM~}

\newcommand{\rvir}{r_{\rm vir}}

\newcommand{\baredth}{\;\overline{\raise1.0pt\hbox{$'$}\hskip-6pt
\partial}\;}
\newcommand{\edth}{\;\raise1.0pt\hbox{$'$}\hskip-6pt\partial\;}

\begin{document}
\title[Three-Point Correlation Function]
{The Three-Point Correlation Function in Cosmology}

\author[M. Takada \& B. Jain]
{Masahiro Takada \thanks{E-mail: mtakada@hep.upenn.edu}
and Bhuvnesh Jain \thanks{E-mail: bjain@physics.upenn.edu} \\
 Department of Physics
and Astronomy, University of Pennsylvania, 209 S. 33rd Street,
Philadelphia, PA 19104, USA
}

\onecolumn
\pagerange{\pageref{firstpage}--\pageref{lastpage}}

\maketitle

\label{firstpage}
\begin{abstract}
With the advent of high-quality surveys in cosmology the 
full three-point correlation function will be a valuable
statistic for describing structure formation models.  
It contains information on cosmological parameters and 
detailed halo properties that cannot be extracted from the 
two-point correlation function. We use the halo 
clustering model to analytically calculate the three-point
correlation function (3PCF) for general cosmological 
fields. We present detailed 
results for the configuration dependence of the 3-dimensional 
mass and galaxy distributions and the 2-dimensional 
cosmic shear field.  We work in real space, where 
higher order correlation functions on small
scales are easier to measure and interpret, but halo model calculations 
get rapidly intractable. Hence we develop techniques
that allow for the accurate evaluation of all the contributing terms
to real space correlations. 
We apply them to the 3PCF to show how its configuration and scale 
dependence changes as one transits from the nonlinear 
to the quasilinear regime, and how this depends on the relative
contributions from the 1-, 2- and 3-halo terms. 

The 3PCF violates the hierarchical ansatz in both its scale and
configuration dependence. We study the behavior of the coefficient $Q$
in the expansion: $\zeta(r_{12},r_{23},r_{31}) =Q[\xi(r_{12})
\xi(r_{23})+\xi(r_{12})\xi(r_{31})+\xi(r_{23})\xi(r_{31})]$ from 
large, quasilinear scales down to about 20 kpc. We find that the nonlinear 3PCF
is sensitive to the halo profile of massive halos, especially its inner
slope. We model the distribution of galaxies in halos and show that the
3PCF of red galaxies has a weaker configuration and scale dependence than the
mass, while for blue galaxies it is very sensitive to the parameters of
the galaxy formation model. The 3PCF from weak lensing on the other 
hand shows different scalings due to projection effects and a sensitivity 
to cosmological parameters. We discuss how our results can be applied 
to various analytical calculations: covariances of
two-point correlation function, the pairwise peculiar velocity dispersion,
higher-order shear correlations, and to extend the halo model by
including the effects of halo triaxiality and substructure on
statistical measures.
\end{abstract}

\section{Introduction}
Understanding the nature of the large-scale structure and the
evolutionary history of the universe is the central aim of
cosmology.  
Therefore, quantifying the observed distribution of matter
rigorously is of fundamental importance, and $n$-point correlation
functions have been the most widely used statistical tools for this
purpose (e.g., see Peebles 1980). 
The inflationary scenario predicts Gaussian initial conditions. 
The statistical properties of the Gaussian field are fully
characterized by the two-point correlation function (2PCF) or its
Fourier transform, the power spectrum.  Based on this idea, the
two-point statistical measures for various cosmological fields have been
extensively used for testing the paradigm as well as constraining
cosmological parameters; the galaxy distribution in the redshift catalog
(e.g., Totsuji \& Kihara 1969; 
Hamilton \& Tegmark 2002), the cosmic microwave background
anisotropy (e.g., de Bernardis 2000) and the cosmic shear field (e.g.,
Van Waerbeke et al. 2001).  
However, even for Gaussian initial conditions,
nonlinear gravitational instability induces non-Gaussian
signatures in the mass distribution, which contain
information on the nature of gravity and the dark matter. 
The three-point correlation
function (3PCF), or equivalently, its Fourier-transformed counterpart,
the bispectrum, is the lowest order statistical tool to probe the
non-Gaussianity. The 3PCF can place strong constraints on models of
structure formation. As an example, the
measurements of the bispectrum in the galaxy distribution at the large
scales ($\simgt 
10~ h^{-1}{\rm Mpc}$)  have been used to determine the bias parameter
as well as to constrain the primordial
non-Gaussianity (e.g., Feldmann et al. 2001; Verde et al. 2001).  It is
also expected that the 3PCF of the cosmic shear field can be used to
precisely determine the cosmological parameters complementary to the
measurements of the 2PCF (e.g., Bernardeau, Mellier \& Van Waerbeke
2002; Bernardeau, Van Waerbeke, \& Mellier 2003).

In this paper, we focus on developing a theoretical model of the 3PCF in
real space for 3D and 2D cosmological fields.  In practice, on small
scales the 3PCF would be easier to measure from observational data over
the bispectrum, since it does not require the Fourier transform for the
survey data that usually have a complicated geometry of data fields.

Theoretical models of the weakly nonlinear 3PCF have been well
studied in the literature based on perturbation theory
\cite{Fry84b,Jing97,Gaz98,Frie99,Barriga02}. Perturbation theory can describe
properties of the dark matter and galaxy clustering on large scales
$\simgt 10~ h^{-1}{\rm Mpc}$ and 
predict that the 3PCF depends on the shape of
triangle configuration and, as a result, contains 
information of the primordial power spectrum and the galaxy
biasing. Historically, the pioneering measurement of the galaxy 3PCF done
by Peebles \& Groth (1975) (also Groth \& Peebles 1977) proposed the
``hierarchical form'',
$\zeta(r_{12},r_{23},r_{31})=Q[\xi(r_{12})\xi(r_{23})
+\xi(r_{23})\xi(r_{31})+\xi(r_{31})\xi(r_{12})], $ with the constant
$Q\simeq 1.3$. However, subsequent work has revealed that
the measured 3PCF does not obey the hierarchical form rigorously and the
large-scale amplitudes can be explained by the perturbation theory
results of the cold dark matter (CDM) models, if the biasing relation is
correctly taken into account for the analysis \cite{Jing98,Frie99}.

On the other hand, a quantitative theoretical model of the 3PCF in the strongly
nonlinear regime is still lacking except for studies relying on 
$N$-body simulations (Matsubara \& Suto 1994; Suto \& Matsubara 1994; Jing
\& B\"orner 1998) \footnote{The nonlinear bispectrum has been well
studied based on extended perturbation theory (Scoccimarro \&
Friemann 1999; also see Bernardeau et al. 2002a) or the recently
developed dark matter halo approach (Ma \& Fry 2000c; Scoccimarro et
al. 2001; Cooray \& Hu 2001a).}. Simulations provide only limited
physical insight into the complex non-linear phenomena involved in 
gravitational clustering, and are intractable for performing multiple
evaluations in parameter space. 

Therefore, the main purpose of this paper is to develop an analytical
model for predicting the 3PCF applicable to both the linear
and nonlinear regimes. 
For this purpose, we
need a model to correctly describe the redshift evolution and
statistical properties of gravitational clustering up to the three-point
level. We employ the so-called dark matter halo model, where
gravitational clustering is described in terms of correlations between
and within dark matter halos.  Originally, this model was developed to
express nonlinear clustering as the real-space
convolution of halo density profiles
(Neymann \& Scott 1952; Peebles 1974; McClelland \& Silk 1977; and also
see Scherrer \& Bertschinger 1991; Sheth \& Jain 1997; Yano \& Gouda
1999; Ma \& Fry
2000b,c). Most recent works have relied on the the Fourier-space formulation, 
since the forms of the power spectrum and the bispectrum become
much simpler (Seljak 2000; Ma \& Fry 2000c; Peacock \& Smith 2000;
Scoccimarro et al. 2001; Cooray \& Hu 2001a,b; Berlind \& Weinberg 2002;
Hamana, Yoshida \& Suto 2002; Takada \& Jain 2002 hereafter TJ02;
Scranton 2002; and also Cooray \& Sheth 2002 for a recent review). 
The halo model appears remarkably
successful in that, even though it relies on rather simplified
assumptions, it has reproduced results from numerical simulations
(Seljak 2000; Ma \& Fry 2000c; Scoccimarro et al. 2001; TJ02)
and also allowed for
interpretations of observational results of galaxy clustering
\cite{Seljak00,Scocci01}. 

We formulate the 3PCF model so that it can be applied to general 3D and 2D
cosmological fields, such as the mass and galaxy distributions and the cosmic
shear fields.  Our method is built on the real-space formulation 
for the correlations of three particles in one halo.  This is because
the real-space approach enables us to compute the one-halo contribution
to the 3PCF by a 4-dimensional integration, which is an 
advantage compared to the
Fourier space approach. For the 2- and 3-halo terms, we rely on the 
Fourier-space approach and the approximations presented in
Scoccimarro et al. (2001; also see TJ02). We study the
transition from the quasi-linear to nonlinear regimes and the 
relative contribution of the different terms to the 3PCF. 
We show the halo model predictions for the 3PCF of
the mass and galaxy distribution and the weak lensing convergence field
for the currently favored CDM model. To do this, we will focus
on the dependences of the 3PCF on the triangle configurations
as well as on the properties of the halo profile, its inner slope and 
concentration.

This paper is organized as follows.  In \S \ref{halo} we present the
formalism to compute the 3PCF of the 3D density field based on the
real-space halo model and review the Fourier-space halo approach and
halo model ingredients (halo density profile, mass function and bias
model) we will use in this paper. In \S \ref{ang3pt}, we present the
method to compute the angular 3PCF.  
In \S \ref{results} we show the results of
the halo model predictions for the 3PCF of the mass and galaxy
distributions and the weak lensing convergence field. Finally, \S
\ref{discuss} is devoted to a summary and discussion. 
We give some useful approximations for calculating the 3PCF in Appendix
\ref{app3pt} and \ref{app}.  
We use the currently favored CDM model ($\Lambda$CDM)
with $\Omega_{m0}=0.3$, $\Omega_{\lambda0}=0.7$, $h=0.7$ and
$\sigma_8=0.9$.  Here $\Omega_{\rm m0}$ and $\Omega_{\lambda0}$ are the
present-day density parameters of matter and cosmological constant, $h$
is the Hubble parameter, and $\sigma_8$ is the rms mass fluctuation in a
sphere of $8h^{-1}$Mpc radius.  The choice of $\sigma_8$ for this model
is motivated by the cluster abundance analysis \cite{Eke96}.

\section{Dark Matter Halo Approach}
\label{halo}

\subsection{Real-space halo approach}
\label{rhalo} 

In this section, we briefly review the {\rm real-space}
dark matter halo approach (McClelland \& Silk
1977; Scherrer \& Bertschinger 1991; Sheth \& Jain 1997; Yano \& Gouda
1999; Ma \& Fry 2000b,c), where the $n$-point correlation functions 
are described by the real-space convolution of halo density profiles.

In the halo approach, we assume that all the matter is in 
halos with the density profile
$\rho_h(\bm{x}; m)$ parameterized by a mass $m$.
 The halo mass is given by
\begin{equation}
m \equiv \int_{V_{\rm vir}}\!\!d^3\bm{x}~ \rho_h(\bm{x}; m),
\label{eqn:massdef}
\end{equation}
where $V_{\rm vir}$ is the virial volume of the halo.  It is convenient
to introduce the normalized halo profile defined as $u_m(\bm{x};
m)=\rho_h(\bm{x}; m)/m$, satisfying the condition
\begin{equation}
\int_{V_{\rm vir}}\!\!d^3\bm{x}~ u_m(\bm{x}; m)=1. 
\label{eqn:nhalop}
\end{equation}
In this paper we employ the {\em virial} volume as the
boundary of a given halo, which can be formally defined,
for example, based on the spherical top-hat collapse model.  However,
since in reality some matter is distributed outside the virial region,
it is non-trivial to choose which boundary condition to use for the halo
model. We will discuss how this uncertainty affects the halo
model predictions, and show that it does not matter in the non-linear
regime.

The density field at some point can be given by the superposition of
each halo density profile: $\rho(\bm{x})=\sum_{i}m_i u_{m_i}(\bm{x};
m_i)$. The mean density of the universe is thus obtained from the
ensemble average (Scherrer \& Bertschinger 1991)
\begin{eqnarray}
\bar{\rho}_0&=&\skaco{\rho(\bm{x})}=\kaco{\sum_{i}m_iu(\bm{x}_i-\bm{x}; m_i)}
=\int\!\!dm~ m n(m) \int\!\!d^3\bm{x}'u_m(\bm{x}-\bm{x}'),
\label{eqn:avedens}
\end{eqnarray}
where we have replaced the ensemble average by a spatial average and
an average over the halo mass function $n(m)$: 
$\skaco{\sum_i\delta_D(m-m_i)\delta_D^3(\bm{x}'-\bm{x}_i)}=n(m)$.  Note
that from equation (\ref{eqn:nhalop}) the integral over $\bm{x}'$ is
unity, so the above equation is a statement about the normalization
of the halo mass function. 
Throughout this paper we work in the comoving coordinate.

Within the framework of the halo approach, the two-point correlation
function (2PCF) of the density field can be expressed as the sum of
correlations within a single halo (1-halo term) and between different
halos (2-halo term):
\begin{eqnarray}
\xi(r)&=&\xi_{1h}(r)+\xi_{2h}(r)\nonumber\\
&=&
\int\!\!dm~ n(m)\left(\frac{m}{\bar{\rho}_0}\right)^2
\int_{V_{\rm vir}}\!\!d^3\bm{x}~ u_m(\bm{x})u_m(\bm{x}+\bm{r})\nonumber\\
&&+\int\!\!dm_1~ n(m_1)\frac{m_1}{\bar{\rho}_0}
\int\!\!dm_2~ n(m_2)\frac{m_2}{\bar{\rho}_0}
\int_{V_{\rm vir}}\!\!d^3\bm{x}_1~ u_{m_1}(\bm{x} - \bm{x}_1)
\int_{V_{\rm vir}}\!\!d^3\bm{x}_2~ u_{m_2}(\bm{x}'-\bm{x}_2)
\xi(\bm{x}_1-\bm{x}_2; m_1,m_2),
\label{eqn:r2ptfull}
\end{eqnarray}
where $\xi(\bm{x}_1-\bm{x}_2; m_1,m_2)$ 
is the 2PCF of two halos with mass $m_1$ and
$m_2$, and we have used $\bm{r}=\bm{x}-\bm{x}'$. 
Here we have again employed the virial volume for
the integration range.  In the other words, the integration cutoff
reflects the fact that we only account for mass contributions out to
the virial region.  If we require that $\xi(r)$ is a function of the
separation $r$ alone (from statistical homogeneity and isotropy), 
then the density profile $u_m$ which we need here is 
an average over all halos of a given mass so that the resulting 2PCF has
no particular direction dependence of $\bm{r}$ (see similar discussion
in Seljak 2000).  This holds for a spherically symmetric density
profile, which is assumed in this paper for simplicity. 
If one considers non-spherically symmetric profiles, it requires
some extra averages over all possible shapes of halos of a given mass.
Equation (\ref{eqn:r2ptfull}) shows that we can obtain $\xi_{1h}(r)$ by
a 3-dimensional integration for a spherically symmetric density
profile, while the 2-halo term needs an 8-dimensional integration at
most.

In the same spirit as the derivation of equation (\ref{eqn:r2ptfull}), we 
can derive  expressions for the 1-halo contribution to the $n$-point 
correlation function ($n\ge 2$): 
\begin{equation}
\xi^{n}_{1h}(\bm{r}_1,\bm{r}_2,\dots,\bm{r}_n)
=\int\!\!dm~ n(m)\left(\frac{m}{\bar{\rho}_0}\right)^n
\int_{V_{\rm vir}}\!\!d^3\bm{x}~ u_m(\bm{x}+\bm{r}_1)u_m(\bm{x}+\bm{r}_2)\cdots
u_m(\bm{x}+\bm{r}_n).
\label{eqn:rnpt} 
\end{equation}
Interestingly, this equation means that we can obtain any $n$-point
correlation function by a 4-dimensional integration, once the density
profile and the mass function are given. The 1-halo contribution is
expected to be dominant in the strongly non-linear regime $(\delta\gg
1)$, and so the real-space halo approach can be useful for calculations
in this regime. 

In this paper, we focus on the halo model predictions of the 3PCF, 
which is the lowest order statistic for describing 
non-Gaussianity of the density field. The 1-halo term can be expressed as
\begin{equation}
\zeta_{1h}(\bm{r}_1,\bm{r}_2,\bm{r}_3)=
\int\!\!dm~ n(m)\left(\frac{m}{\bar{\rho}_0}\right)^3
\int_{V_{\rm vir}}\!\!d^3\bm{x}~ u_m(\bm{x}+\bm{r}_1)
u_m(\bm{x}+\bm{r}_2)u_m(\bm{x}+\bm{r}_3).
\label{eqn:r3pt}
\end{equation}
Under the assumption of statistical symmetry, the 3PCF can be described by
three independent parameters characterizing triangle configurations.
Likewise, the 2- and 3-halo terms in 3PCF can be derived within the
real-space halo approach (Ma \& Fry 2000c); for example, the 2-halo term
becomes
\begin{eqnarray}
\zeta_{2h}(\bm{r}_1,\bm{r}_2,\bm{r}_3)
&=&\int\!\!dm~ n(m)\left(\frac{m}{\bar{\rho}_0}\right)^2 \int\!\!dm'~ 
n(m')\frac{m'}{\bar{\rho}_0}\nonumber\\
&&\times 
\int_{\rm V_{\rm vir}}\!\!d^3\bm{x}~ u_m(\bm{x}+\bm{r}_1)u_m(\bm{x}+\bm{r}_2)
\int_{\rm V^\prime_{\rm vir}}\!\!d^3\bm{x}'~ u_{m'}(\bm{x}'+\bm{r}_3)
\xi(|\bm{x}-\bm{x}'|;m, m') + \mbox{perm}(1,2,3).
\end{eqnarray}
This equation implies that we have to perform at most an 8-dimensional
integration to get $\zeta_{2h}$ for a given halo-halo correlation
function, $\xi(r; m,m')$, while the 3-halo contribution to $\zeta$
similarly requires the 12-dimensional integration.

\subsection{Fourier-space halo approach}
\label{fhalo}

Most recent studies of the halo approach has relied on the method build
in the Fourier space (Seljak 2000; Ma \& Fry 2000c; Peacock \& Smith
2000; Scoccimarro et
al. 2001; Cooray \& Hu 2001a,b; TJ02), since the
Fourier-transformed counterparts of the $n$-point correlation function
(power spectrum, bispectrum and etc.) can be expressed simply. 

Following the notation introduced in Cooray \& Hu (2001a), the power
spectrum can be expressed as the sum of the 1-halo and 2-halo
contributions
\begin{equation}
P(k)=P_{1h}(k)+P_{2h}(k)=I^0_2(k,k)+\left[I^1_1(k)\right]^2P^L(k), 
\label{eqn:ps}
\end{equation}
with 
\begin{equation}
I^\beta_\mu(k_1,\dots,k_{\mu})\equiv \int\!\!dm~ n(m)
b^\beta(m)\left(\frac{m}{\bar{\rho}_0}\right)^\mu 
\tilde{u}_m(k_1)\dots \tilde{u}_m(k_\mu).
\end{equation}
Here $\tilde{u}_m(k)$ is the Fourier transform of the density profile
and the following definition has been used recently
(e.g., Seljak 2000; Scoccimarro et al. 2001; Cooray \& Sheth 2002):
\begin{equation}
\tilde{u}_m(k)=\int_0^{\rvir}\!4\pi r^2dr~ u_m(r)j_0(kr),
\label{eqn:fum}
\end{equation}
where $j_0$ is the zeroth-order spherical Bessel function, $j_0(x)=\sin
x/x$, and $r_{\rm vir}$ denotes the virial radius. 
In the following, quantities with
tilde symbols denote the corresponding Fourier transformed quantities.
Notice that all relevant quantities also depend on redshift, although we
often omit it in the argument for simplicity.  
In contrast to the truncated halo profile above,
Ma \& Fry (2000c) employed the non-truncated profile whose Fourier
transform is defined over an infinity integration range as
 $\tilde{u}_m(k)=\int^\infty_0\!\! 4\pi
r^2dru_m(r)j_0(kr)$.  The different halo boundary 
leads to different results in the Fourier-space halo model.
In particular,  we will focus on the difference between the predictions for 
the mass bispectrum in Ma
\& Fry (2000b), Scoccimarro et al. (2001) and Cooray \& Sheth (2002) in
comparison with the real-space halo model for the 3PCF in this paper.
In the derivation of equation (\ref{eqn:ps}), we have assumed that the
power spectrum for the halo-halo 2PCF can be given by $P(k;
m_1,m_2)=b(m_1)b(m_2)P^L(k)$, where $P^L(k)$ is the linear power
spectrum and $b(m)$ denotes the bias parameter between the distribution
of halos and the underlying density field.  Through the definition
$\xi(r)=\int_0^\infty\!\!k^2dk/(2\pi^2)P(k)j_0(kr)$, we can compute the
2PCF of the density field from the power spectrum (\ref{eqn:ps}) within
the framework of the Fourier-space halo model. The two derivations of
$\xi(r)$ based on the real-space and Fourier-space halo approaches
should be equivalent to each other.  However, the relation between
$\xi(r)$ and $P(k)$ uses a Fourier transform over the infinite volume,
while the integrations used in the halo model (see equations
(\ref{eqn:r2ptfull}) and (\ref{eqn:fum})) are confined to the virial
volume of halos. This could lead to some discrepancies between the two
halo model predictions for $\xi(r)$, which will be carefully
investigated below.

Likewise, the Fourier-transformed counterpart of the mass 3PCF, the
bispectrum, can be expressed as the sum of the three contributions where
three dark matter particles reside in one halo (1-halo) or in
two (2-halo) and three (3-halo) different halos, respectively:
\begin{eqnarray}
B(\bm{k}_1,\bm{k}_2,\bm{k}_3)&=&I^0_3(k_1,k_2,k_3)
+\left[I_2^1(k_1,k_2)I_1^1(k_3)P^L(k_3)+\mbox{perm}(1,2,3)\right]
+I_1^1(k_1)I_1^1(k_2)I^1_1(k_3)B^{PT}(\bm{k}_1,\bm{k}_2,\bm{k}_3).
\label{eqn:bisp}
\end{eqnarray}
Here $B^{PT}$ denotes the perturbation theory bispectrum (e.g., 
see Jain \& Bertschinger 1994) given by
\begin{equation}
B^{PT}(\bm{k}_1,\bm{k}_2,\bm{k}_3)=\left[\frac{10}{7}+\left(\frac{k_1}{k_2}
+\frac{k_2}{k_1}\right)\frac{\bm{k}_1\cdot\bm{k}_2}{k_1k_2}+\frac{4}{7}
\frac{(\bm{k}_1\cdot\bm{k}_2)^2}{k_1^2k_2^2}\right]P^L(k_1)P^L(k_2)
+\mbox{perm}. 
\end{equation}
Here we have neglected the weak dependences of $B^{PT}$ on the
cosmological parameter $\Omega_{\rm m0}$.
Using the bispectrum, the 3PCF can be defined as
\begin{equation}
\zeta(\bm{r}_1,\bm{r}_2,\bm{r}_3)=\int\!\!\prod_{i=1}^3
\frac{d^3\bm{k}_i}{(2\pi)^3} B(\bm{k}_1,\bm{k}_2,\bm{k}_3)
\exp[i(\bm{k}_1\cdot\bm{r}_1+\bm{k}_2\cdot\bm{r}_2+\bm{k}_3\cdot\bm{r}_3)]
(2\pi)^3\delta_D(\bm{k}_{123}),
\label{eqn:f3pt}
\end{equation}
where $\bm{k}_{123}=\bm{k}_1+\bm{k}_2+\bm{k}_3$. 

\subsection{Computational ease in real-space versus Fourier-space}

Before specifying the details of our method, we briefly 
compare the real-space and Fourier-space halo approaches for
predicting the 3PCF of the mass distribution.

Let us first consider the 1-halo contribution to the 3PCF.  Equation
(\ref{eqn:r3pt}) shows that, once the halo profile and mass function are
given, the real-space halo model enables us to obtain the 1-halo term by
a 4-dimensional integration for arbitrary triangle shapes.  On the
other hand, equations (\ref{eqn:bisp}) and (\ref{eqn:f3pt}) imply that
a 7-dimensional integration at most is required to get the 1-halo term
based on the Fourier-space halo model, which is intractable with current
computational expenses. Therefore, the real-space halo approach has
a great advantage in predicting the 1-halo term of the 3PCF. 

However, the Fourier-space approach becomes useful for predicting
the 2-halo and 3-halo terms, because the real-space model requires 
at most 8- and 12-dimensional integrations for calculating the 2-
and 3-halo terms for given 2PCF and 3PCF of the halo distribution. On
the other hand, from equations (\ref{eqn:bisp}) and (\ref{eqn:f3pt}),
one can see that the Fourier-space model allows us to obtain the 2- and
3-halo terms basically by a 7-dimensional integration. However, 
direct integration is still intractable.  Therefore, in this paper we
employ the Fourier-space model and develop approximations for
calculating the 2- and 3-halo terms.  Those approximations can
significantly reduce the computational expenses and make the calculations
tractable by regular numerical integrations, but 
still have adequate accuracy for our purpose.

\subsection{Ingredients of the halo model}
\label{ingred}

To complete the halo model approach, we need suitable models for the three
ingredients: the halo density
profile, the mass function of halos and the biasing of the halo distribution,
each of which depends on halo mass $m$ and redshift
$z$ and have been well studied in the literature. 

We consider density profiles of the form
\begin{equation}
\rho_h(r)=\frac{\rho_s}{(cr/\rvir)^\alpha (1+cr/\rvir)^{3-\alpha}},
\label{eqn:nfw}
\end{equation}
where $\rho_s$ is the central density parameter and $c$ is the
concentration parameter. In most parts of this paper, we will take the 
NFW profile with $\alpha=1$ (Navarro, Frenk \& White 1997). However,
since other
simulations with higher spatial resolution have indicated
$\alpha\approx -1.5$ \cite{Fukushige97,Moore98,JS00}, we will
consider the effects of the variation in $\alpha$ on the 2- and 3-point 
correlation functions. 
The parameter $\rho_s$ can be eliminated by the 
definition (\ref{eqn:massdef}) of halo mass as
\begin{eqnarray}
m=\int_0^{\rvir}\!\!4\pi r^2dr \rho_h(r)
=\frac{4\pi \rho_s \rvir^3}{c^3}\times 
\left\{
\begin{array}{ll}
\displaystyle f^{-1}, & (\mbox{NFW}; \alpha=1),\\
\displaystyle 
\frac{c^{3-\alpha}}{3-\alpha}~ 
{}_2F_1(3-\alpha,3-\alpha,4-\alpha,-c), & (\mbox{otherwise}), 
\end{array}
\right.
\end{eqnarray}
where $f =1/[ \ln(1+c)-c/(1+c)]$ and ${}_2F_1$ denotes the
hypergeometric function. We employ the virial radius given by the
spherical top-hat collapse model: $m=(4\pi \rvir^3/3)
\bar{\rho}_0\Delta(z)$, where $\Delta(z)$ is the overdensity of collapse
given as a function of redshift (e.g., see Nakamura \& Suto 1997 and
Henry 2000 for a useful fitting formula). We have $\Delta\approx 340$
for the \LCDM model.

To give the halo profile in terms of $m$ and $z$, we further need to
express the concentration parameter $c$ in terms of $m$ and $z$;
however, this still remains somewhat uncertain.  We consider the
following $c$ parameterized by the normalization and the mass slope: 
\begin{equation}
c(m,z)=c_0(1+z)^{-1}\left(\frac{m}{m_\ast(z=0)}\right)^{-\beta},
\label{eqn:conc}
\end{equation}
where $m_\ast(z=0)$ is the nonlinear mass scale defined as
$\delta_c(z=0)/D(z=0)\sigma(m_\ast)=1$. Here $\delta_c$ is the threshold
overdensity for spherical collapse model (see Nakamura \& Suto 1997 and
Henry 2000 for a useful fitting function), $\sigma(m)$ is the present-day
rms fluctuations in the matter density top-hat smoothed over a scale
$R_m\equiv (3m/4\pi\bar{\rho}_0)^{1/3}$, and $D(z)$ is the growth
factor (e.g., see Peebles 1980).  The values $c_0\sim 10$ and $\beta\sim
O(10^{-1})$ are theoretically expected.  The redshift dependence
$(1+z)^{-1}$ is taken based on the numerical simulation results in
Bullock et al. (2001). We take $c_0=10$ and $\beta=0.2$ as our reference
model for the NFW profile, because TJ02 pointed out that the model can
reproduce simulation results for the variance,
skewness of weak lensing fields and the shear kurtosis. We will consider
the influence of variations in the concentration parameter on the halo
model predictions of the 2PCF and 3PCF.

We will often use the following useful form for the Fourier
transform of the NFW profile (Scoccimarro et al. 2001), which is
needed for the Fourier-space halo model calculations:
\begin{equation}
\tilde{u}_m(k)=f\left[\sin\eta\left\{
{\rm Si}(\eta(1+c))-{\rm Si}(\eta)\right\}
+\cos\eta\left\{{\rm Ci}(\eta(1+c))-{\rm Ci}(\eta)\right\}
-\frac{\sin(\eta c)}{\eta(1+c)}\right],
\label{eqn:fumnfw}
\end{equation}
where $\eta\equiv k \rvir/c$, ${\rm Ci}(x)=-\int^\infty_xdt\cos t/t$ is the
cosine integral function and ${\rm Si}(x)=\int^x_0\!\!dt\sin(t)/t$ the
sine integral. The profile (\ref{eqn:nfw}) with general $\alpha$ has the
asymptotic behaviors of $\tilde{u}_m(k) \approx 1$ and $\propto k^{-3+\alpha}$
for $k\ll c/\rvir $ and $k\gg c/\rvir$, respectively. 
It is worth stressing that
the equation above is derived from the NFW profile truncated at the
virial radius to maintain mass conservation (see equations (\ref{eqn:massdef})
and (\ref{eqn:avedens})). The truncated profile leads to 
the consequence that the 1-halo term of
the power spectrum behaves like shot-noise as $P_{1h}(k)\propto k^0$
at small $k$.  
If one employs the non-truncated NFW profile,  
the resulting $\tilde{u}_m(k)$ is
logarithmically divergent at $k\rightarrow 0$ (see equation (22) in Ma
\& Fry 2000b) originating from the mass divergence from $\int^\infty_0\!
\! 4\pi r^2 dr\rho_h(r)$, which does not yield the shot-noise power 
spectrum. 

For the halo mass function, we adopt an analytical fitting model
proposed by Sheth \& Tormen (1999), which is more accurate on cluster
mass scales than the original Press-Schechter model \cite{PS74}. The
number density of halos with mass in the range between $m$ and $m+dm$ is
given by
\begin{eqnarray}
n(m)dm&=&\frac{\bar{\rho}_0}{m}f(\nu)d\nu\nonumber\\
&=&\frac{\bar{\rho}_0}{m}A[1+(a\nu)^{-p}]\sqrt{a\nu}\exp\left(-\frac{a\nu}{2}
\right)\frac{d\nu}{\nu},
\label{eqn:massfun}
\end{eqnarray}
where $\nu$ is the peak height defined by
\begin{equation}
\nu=\left[\frac{\delta_c(z)}{D(z)\sigma(m)}\right]^2,
\end{equation}
and the numerical coefficients $a$ and $p$ are empirically fitted from
$N$-body simulations as $a=0.707$ and $p=0.3$. The coefficient $A$ is set
by the normalization condition $\int_0^\infty\!d\nu f(\nu)=1$, leading
to $A\approx 0.129$. Note that the peak hight $\nu$ is specified as a
function of $m$ at any redshift once the cosmological model is fixed.

Mo \& White (1996) developed a useful formula to describe the bias
relation between the halo distribution and the underlying
density field.  This idea has been improved by several authors using
$N$-body numerical simulations \cite{Mo97,Jingbias98,Sheth99,ST99}; 
we will use the
fitting formula of Sheth \& Tormen (1999) for consistency with the mass
function (\ref{eqn:massfun}):
\begin{equation}
b(\nu)=1+\frac{a\nu-1}{\delta_c}+\frac{2p}{\delta_c(1+(a\nu)^p)},
\label{eqn:bias}
\end{equation}
where we have assumed scale-independent bias and neglected the higher
order bias functions $(b_2, b_3,\cdots)$. This bias model is used for
calculations of the 2-halo term in the 2PCF and the 2- and 3-halo terms in
the 3PCF (see equations (\ref{eqn:ps}) and (\ref{eqn:bisp})). It should be
noted that the requirement that the 2-halo term in the power spectrum
gives the linear power spectrum in the limit $k\rightarrow 0$
($\tilde{u}_m(k)\rightarrow 1$) imposes the condition $\int\!\!d\nu
f(\nu)b(\nu)=1$ for the integration mass range, since we have assumed
that all the matter is in the form of virialized halos.

\section{Angular $n$-point correlation function}
\label{ang3pt}

In this section, we present the angular
$n$-point correlation functions in analogy with the real-space halo
approach to the 3D $n$-point correlation functions in \S
\ref{rhalo}. The results are applicable for a general projected cosmological
field such as the angular galaxy distribution and weak lensing
fields.

\subsection{Projected density field}
We first consider the 2-dimensional projection
$\Sigma(\bm{\theta})$ of the density field 
along the line of sight:
\begin{equation}
\Sigma(\bm{\theta})=\int\!\!d\chi~ W(\chi)\ \rho(\chi,d_A\bm{\theta}), 
\label{eqn:2dfield}
\end{equation} 
where $W(\chi)$ is the weight function, and $\chi$ and $d_A(\chi)$ are
the comoving distance and comoving angular diameter distance, respectively.
Note that $\chi$ is related to redshift $z$ via the relation
$d\chi=dz/H(z)$. In this paper we take the projected field to 
be the weak lensing convergence field $\kappa$ (e.g., 
see Bartelmann \& Schneider 2001) with the weight function given by
\begin{eqnarray}
W(\chi,\chi_s)=\frac{3}{2}\Omega_{m0}H_0^2a^{-1}
\frac{d_A(\chi)d_A(\chi_{\rm s}-\chi)}{d_A(\chi_s)},
\label{eqn:wgl} 
\end{eqnarray}
where we have considered a single source redshift $z_{s}$, 
corresponding to the comoving distance $\chi_s$, for simplicity. 
The expressions derived below can be applied to the angular clustering 
of galaxies
as well. Setting $W(\chi)=dN/d\chi$, the selection function of 
galaxies, and replacing the density field $\rho$ by $\rho_{\rm galaxy}$
(see \S 5 below for explicit expressions in terms of the halo
occupation number of galaxies) will give the projected density
of galaxies $\Sigma_{\rm galaxy}$ from equation (\ref{eqn:2dfield}),
and can be used to replace $\Sigma_m$ in the equations below. 

\subsection{The  1-halo contribution to 
angular $n$-point correlation functions}

Analogous to the 3D real-space halo approach, we can
express the 1-halo contribution to the angular 
2-point correlation function of the convergence field as
\begin{eqnarray}
w_{1h}(\theta)=\int^{\chi_s}_0\!\!d\chi \frac{d^2V}{d\chi d\Omega}
W^2(\chi) \int\!\!dm~
n(m; \chi)\left(\frac{m}{\bar{\rho}_0}\right)^2 
\int_{\Omega_{\rm vir}}\!\!d^2\bm{\varphi}~ \Sigma_m(\bm{\varphi}; \chi)
\Sigma_{m}(\bm{\varphi}+\bm{\theta}; \chi),
\label{eqn:r2d2pt}
\end{eqnarray}
where $d^2V/d\chi d\Omega$ is the comoving differential volume given by
$d^2V/d\chi d\Omega=d_A^2(\chi)$ for a flat universe 
and we have assumed a circularly
symmetric profile $\Sigma_m$ for halo of a given mass.  We have also
employed the flat-sky approximation, the Limber approximation (Kaiser
1992), which means that we account for contributions to $w_{1h}$ from
lens structures at the {\em same} redshift, and the Born approximation.
  Finally, as in equation (\ref{eqn:r2ptfull}),
we have restricted the integration range to the circular area 
$\Omega_{\rm vir}$ enclosed by the
projected virial radius.  Equation
(\ref{eqn:r2d2pt}) implies that we can obtain $w_{1h}$ by a
4-dimensional integration if we have an analytical
expression for $\Sigma_{m}$.  In the halo model,
$\Sigma_m$ is the normalized column density field for a halo with mass
$m$ defined as
\begin{equation}
\Sigma_m(\bm{\theta})\equiv 
\int_{-\rvir}^{\rvir}\!\!dr_{\parallel} u_m(r_\parallel,d_A\bm{\theta}).
\label{eqn:sigma}
\end{equation}
The factor $d^2V/d\chi d\Omega=d_A^2(\chi)$ in equation
(\ref{eqn:r2d2pt}) arises from the substitution
$d^2\bm{x}_\perp=d_A^2 d^2\bm{\varphi}$, if one wishes to relate
equation  (\ref{eqn:r2d2pt}) to equation (\ref{eqn:r2ptfull}) for the
3-dimensional case. 
It is worth noting that, in analogy with equation
(\ref{eqn:nhalop}), $\Sigma_m$ should be defined so that it satisfies
the mass conservation condition inside the virial radius:
\begin{equation}
\int_{S_{\rm vir}}\!\!d^2\bm{r}_\perp \Sigma_m(\bm{\theta}; \chi)=d_A^2(\chi)
\int_{\Omega_{\rm vir}}\!\!d^2\bm{\theta}~ \Sigma_m(\bm{\theta})=1.
\end{equation}
For the NFW profile (\ref{eqn:nfw}), $\Sigma_m(\theta)$ can be
analytically derived using formulae (2.266) and (2.269.2) 
in Gradshteyn \& Ryzhik (2000) as 
\begin{equation}
\Sigma_m(\theta)=\frac{f c^2}{2\pi \rvir^2}
G(c\theta/\theta_{\rm vir}),
\label{eqn:formsig}
\end{equation}
with
\begin{eqnarray}
G(x)=
\left\{
\begin{array}{ll}
\displaystyle
-\frac{\sqrt{c^2-x^2}}{(1-x^2)(1+c)}+
\frac{1}{(1-x^2)^{3/2}}{\rm arccosh}\frac{x^2+c}{x(1+c)},
& (x<1)\\
\displaystyle \frac{\sqrt{c^2-1}}{3(1+c)}\left[1+\frac{1}{c+1}\right],
& (x=1)\\
\displaystyle
-\frac{\sqrt{c^2-x^2}}{(1-x^2)(1+c)}-\frac{1}{(x^2-1)^{3/2}}
{\rm arccos}\frac{x^2+c}{x(1+c)},
& (1<x\le c)
\end{array} 
\right.
\end{eqnarray}
where $\theta_{\rm vir}$ is the angular size of the virial radius of
a halo at a given distance $\chi$: $\theta_{\rm
vir}=\rvir/d_A(\chi)$. Note that our expression for $\Sigma_m$ 
differs from the result of Bartelmann (1996) 
because he took the projection over $r_\parallel=[-\infty,\infty]$
(his expression includes typos, so,
e.g., see Wright \& Brainerd 2000 for the correct expression). 
The difference turns out to be pronounced, e.g. using the expression for 
$c \rightarrow \infty$ in the above equation leads to discrepancies
between the real- and Fourier-space halo model calculations of
$w_{1h}(\theta)$, as shown below.

It is useful to replace the projected halo profile $\Sigma_m$ in 
equation (\ref{eqn:r2d2pt}) by the convergence field,
$\kappa_m(\bm{\theta})$,  for a halo of mass $m$:
\begin{equation}
\kappa_m(\bm{\theta}; \chi)\equiv W(\chi,\chi_s)\left(\frac{m}{\bar{\rho}_0}
\right)\Sigma_m(\bm{\theta}; \chi)=4\pi G a^{-1}
\frac{d_A(\chi)d_A(\chi_s-\chi)}{d_A(\chi_s)}\int^{\rvir}_{-\rvir}
\!\!dr_{\parallel}~
\rho_h(r_{\parallel},d_A(\chi)\bm{\theta}; m),
\label{eqn:kappam}
\end{equation}
where we have used $(3/2)\Omega_{\rm m0}H_0^2/\bar{\rho}_0 =4\pi G$.  
Hence, equation
(\ref{eqn:r2d2pt}) can be rewritten in a more physically transparent form
\begin{equation}
w_{1h}(\theta)=\int^{\chi_s}_{0}\!\!d\chi~ 
\frac{d^2V}{d\chi d\Omega}\int\!\!dm~ n(m; \chi)
\int_{\Omega_{\rm vir}}\!\!d^2\bm{\varphi} ~\kappa_m(\bm{\varphi}; \chi)
\kappa_m(\bm{\varphi}+\bm{\theta}; \chi). 
\label{eqn:2ptconv}
\end{equation}
This equation means that the 2PCF of the weak lensing convergence can be
expressed as the line-of-sight integration
of the lensing contribution of halos at different 
redshift, weighted with the halo number density. 
An interesting application of this formulation is the study of the 2PCF
of the reduced shear field, $\bm{g}=\bm{\gamma}/(1-\kappa)$
($\bm{\gamma}$ is the shear due to lensing) which is a direct observable
of the cosmic shear measurement.  Equation (\ref{eqn:2ptconv}) no longer
relies on the power spectrum to compute the 2PCF.  Therefore, replacing
$\kappa_m$ in the above equation with $\bm{g}_m$ for a given halo will
yield predictions of the reduced shear 2PCF.  So far cosmological
interpretations of cosmic shear data have been made by comparing the
data with the theoretical model of the shear 2PCF computed from a model
of the 3D power spectrum (e.g., see Van Waerbeke et al. 2001).  The
non-linear correction in using the reduced shear rather than
$\bm{\gamma}$ could be important on sub-arcminute scales, where $\kappa$
is of $O(10^{-1})$ for massive halos.  This study will be presented
elsewhere.

Performing the 2-dimensional Fourier transform of equation (\ref{eqn:2ptconv})
yields the 1-halo contribution to the angular power spectrum of the 
convergence field, $C_{\kappa }(l)$: 
\begin{eqnarray}
C_{\kappa,1h}(l)&=&\int\!\!d\chi \frac{d^2V}{d\chi d\Omega}
\int\!\!dm~ n(m; \chi)
|\tilde{\kappa}_{m}(l)|^2, 
\label{eqn:rcl}
\end{eqnarray}
where we have assumed that the Fourier transform confined to the virial 
region is well approximated as the 2D Fourier transform over an
infinite integration range and 
$\tilde{\kappa}_m(l)$ is the 2D Fourier transform of $\kappa_m$ 
defined as
\begin{equation}
\tilde{\kappa}_{m}(l)\equiv \int^{\theta_{\rm vir}}_0\!2\pi\theta d\theta
~ \kappa_m(\theta)J_0(\theta l).
\end{equation}
Here $J_0(x)$ is the zeroth-order Bessel function.
The expression
(\ref{eqn:rcl}) coincides with the form 
in Cooray et al. (2000) (see also the original 
formulation in Cole \& Kaiser 1988). 

To derive $C_{\kappa}(l)$, the usual approach is to 
model the 3D power spectrum and then employ Limber's
equation (e.g., see TJ02 and see references therein) to get
\begin{equation}
C_{1h}(l)=\int\!\!d\chi W^2(\chi)d_A^{-2}(\chi) 
P\!\!\left(k=\frac{l}{d_A(\chi)}; \chi\right). 
\label{eqn:cl}
\end{equation}
Using the expression (\ref{eqn:ps}) of $P(k)$ based on the Fourier-space
halo model leads to the another expression of $w_{1h}(\theta)$: 
\begin{equation}
w_{1h}(\theta)=\int\!\!d\chi W^2(\chi)d_A^{-2}(\chi)
\int\!\!dm~ n(m;\chi)\left(\frac{m}{\bar{\rho}_0}\right)^2 
\int\!\!\frac{ldl}{2\pi} |\tilde{u}_m(k)|^2 J_0(\theta l),
\label{eqn:2d2pt}
\end{equation}
where $k=l/d_A(\chi)$.  The difference between the real-space and
Fourier-space approaches in equations 
(\ref{eqn:r2d2pt}) and (\ref{eqn:2d2pt}) is
only in the order of the projection. Therefore, the two approaches
should be equivalent to each other (see also discussions in
Cooray et al. 2000; Cooray \& Hu 2001a), if the weight function is a
smooth function with respect to redshift as required for Limber's
approximation.  We will check this by comparing 
the predictions from the two methods 

In a similar manner, we can derive the 1-halo term for
the angular $n$-point correlation
function based on the real-space halo approach as
\begin{eqnarray}
w^n_{1h}(\bm{\theta}_1,\bm{\theta}_2,\dots,\bm{\theta}_n)
=\int\!\!d\chi W^n(\chi) \int\!\!dm~ n(m; \chi)\left(\frac{m}{\bar{\rho}_0}
\right)^n\int_{\Omega_{\rm vir}}\!\!d^2\bm{\varphi}~
\Sigma_m(\bm{\varphi}+\bm{\theta}_1; \chi)\Sigma_m(\bm{\varphi}+\bm{\theta}_2; \chi)
\cdots\Sigma_m(\bm{\varphi}+\bm{\theta}_n; \chi). 
\end{eqnarray}
This equation implies that we can obtain the 1-halo term of any
$n$-point angular correlation functions by a $4$-dimensional
integration.  This expression will be useful in
investigating statistical properties of the projected field at
non-linear angular scales. In this paper, we focus on the 3-point
correlation function of the weak lensing convergence, given by
\begin{eqnarray}
Z_{1h}(\bm{\theta}_1,\bm{\theta}_2,\bm{\theta}_3)&=&
\int^{\chi_s}_0\!\!d\chi~ d_A^2(\chi)
W^3(\chi) \int\!\!dm~ n(m; \chi)\left(\frac{m}{\bar{\rho}_0}
\right)^3\int_{\Omega_{\rm vir}}\!\!d^2\bm{\varphi}~
\Sigma_m(\bm{\varphi}+\bm{\theta}_1;\chi)\Sigma_m(\bm{\varphi}+\bm{\theta}_2; \chi)
\Sigma_m(\bm{\varphi}+\bm{\theta}_3; \chi)\nonumber\\
&=&\int^{\chi_s}_0\!\!d\chi~ d_A^2(\chi)\int\!\!dm~ n(m; \chi)
\int_{\Omega_{\rm vir}}\!\!d\bm{\varphi}~ \kappa_m(\bm{\varphi}+\bm{\theta}_1;\chi)
\kappa_m(\bm{\varphi}+\bm{\theta}_2)\kappa_m(\bm{\varphi}+\bm{\theta}_3; \chi). 
\label{eqn:r2d3pt}
\end{eqnarray}
From statistical symmetry, $Z$ can be given as a function of three
independent parameters characterizing the triangle shape on the sky.

On the other hand, in the conventional approach to
the 3PCF, $Z$ 
is expressed in terms of the 3-dimensional bispectrum as
\begin{equation}
Z(\bm{\theta}_1,\bm{\theta}_2,\bm{\theta}_3)=\int^{\chi_s}_0\!\!d\chi 
W^3(\chi)d_A^{-4}
\int\!\!\prod_{i=1}^{3}\frac{d^2\bm{l}_i}{(2\pi)^2}
B(\bm{k}_1,\bm{k}_2,\bm{k}_3; \chi)\exp[i\bm{l}_i\cdot\bm{\theta}_i
](2\pi)^2\delta_D^2(\bm{l}_{123}),
\label{eqn:f2d3pt}
\end{equation}
where $k_i=l_i/d_A(\chi)$.
We use this formulation 
for evaluating the 2-halo and 3-halo 
contributions to $Z$, using the expressions (\ref{eqn:bisp}) for the 
2- and 3-halo bispectrum. The resulting 
6-dimensional integrations are intractable, so we 
develop approximations, presented in Appendix \ref{app3pt}, to evaluate
them.

\section{Results}
\label{results}
In this section, we present halo model predictions for the three-point
correlation function for
the mass and galaxy distribution and for the weak lensing field. 
We employ a scale invariant spectrum of primordial fluctuations with 
the BBKS transfer function \cite{BBKS86}. 

\subsection{2-point correlation function}

Before considering the 3PCF, we demonstrate the validity
of the real-space halo model by comparing our results for the two-point
correlation function with those of the Fourier-space halo model well
studied in the literature.

For the NFW profile, the 1-halo term in equation (\ref{eqn:r2ptfull})
can be further analytically simplified as
\begin{eqnarray}
\xi_{1h}(r)=\frac{1}{8\pi}
\int\!\!dm~ n(m)\left(\frac{m}{\bar{\rho}_0}\right)^2
\left(\frac{f c^2}{\rvir^2}\right)^2
\int^{\rvir}_0\!\! ds~ (1+c s/\rvir)^{-2}{\cal I}(s; r),
\label{eqn:2ptrcalc}
\end{eqnarray}
with 
\begin{equation}
{\cal I}(s; r)=
\left\{
\begin{array}{ll}
0, & (r\ge 2\rvir \mbox{, or } |r-s|> \rvir)\\
\displaystyle \frac{\rvir-|r-s|}{r(1+c|r-s|/\rvir)(1+c)},& (s+r\ge \rvir)\\
\displaystyle \frac{(r+s)-|r-s|}{r(1+c|r-s|/\rvir)(1+c(r+s)/\rvir)},& 
({\rm otherwise}),
\end{array}
\right.
\end{equation}
where we have taken the virial volume for the
integration range of $d^3\bm{s}$ in equation (\ref{eqn:r2ptfull}) and
the boundary condition $u_m(x)=0$ for $x>r_{\rm vir}$.  The predictions
for $\xi_{1h}$ on nonlinear scales are not sensitive to the boundary
condition, as shown below.
Note that, although the integration over $ds$ can be further analytically 
calculated as done in Ma \& Fry (2000c) and in Sheth et al. (2001), 
the resulting  expression for $\xi_{1h}$
is lengthy for the boundary condition 
and so we stopped at the above expression.

\begin{figure}
  \begin{center}
    \leavevmode\epsfxsize=8.4cm \epsfbox{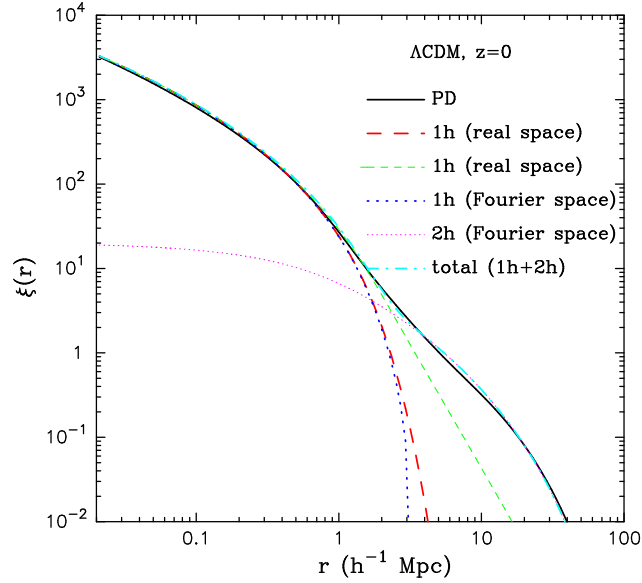}
  \end{center}
\caption{The 2-point correlation function of the mass density field,
$\xi(r)$, at $z=0$ for $\Lambda$CDM. The solid curve is the prediction
from the Peacock \& Dodds fitting formula. The thick and thin dashed
curves show the 1-halo contribution to $\xi(r)$, calculated using the
real-space halo approach with different boundary conditions.
The thick and thin dotted curves
are the 1-halo and 2-halo terms from the Fourier-space halo approach.
The dot-dashed curve is the total prediction of the 1-halo plus 2-halo
terms, where we have used the real-space (thick dashed curve) and
Fourier-space (thin dotted curve) models for the 1-halo and 2-halo
predictions, respectively.  } \label{fig:comp_2pt}
\end{figure}
  Figure \ref{fig:comp_2pt} plots the mass two-point correlation
function $\xi(r)$ at $z=0$ for
the \LCDM model.  The halo model prediction is shown by
the dot-dashed curve, which matches the solid
curve given by the fitting formula of Peacock \&
Dodds (1996; hereafter PD).  
We consider the NFW profile with 
the concentration specified by $(c_0,\beta)=(10,0.2)$ in equation
(\ref{eqn:conc}) as the fiducial model.
During the preparation of this paper, Smith et al. (2002) have proposed
a new fitting formula of the non-linear power spectrum that can better
match high-resolution $N$-body simulations. For the \LCDM
model, the effect is small: the power spectrum differs from 
the PD prediction by $\sim 10\%$ at most at $k\simlt 30 h/$Mpc. 
The thick dashed curve shows the
real-space halo model prediction for the 1-halo contribution to $\xi$
computed from equation (\ref{eqn:2ptrcalc}), while the thick dotted curve is
the prediction from the Fourier-space halo model. The two halo model 
predictions are in remarkable agreement for
small scales $r\simlt 2~ h^{-1}{\rm Mpc}$,  
but deviate slightly at
the transition scale of $\sim 2~ h^{-1}{\rm Mpc}$ between the non-linear
and linear regimes. The discrepancy between the real-space and
Fourier-space halo models could be ascribed to the boundary
condition used in the integration. The thin dashed curve shows the result
obtained without using the boundary condition ($u_m(x)=0$ for $x>\rvir$)
in equation (\ref{eqn:2ptrcalc}), implying that the change alters
the predictions only on scales $r\simgt 1~h^{-1}{\rm Mpc}$.  
However on these scales the 2-halo term dominates
as shown by the thin dotted
curve. Hence, to summarize the results in this figure, the
real-space halo model can be used to predict the statistical properties of
the non-linear density field with the same accuracy
as the Fourier-space halo model whose validity has been carefully
investigated in the literature (Seljak 2000; Ma \& Fry 2000c;
Scoccimarro et al. 2001; TJ02).

\begin{figure}
  \begin{center}
    \leavevmode\epsfxsize=16.cm \epsfbox{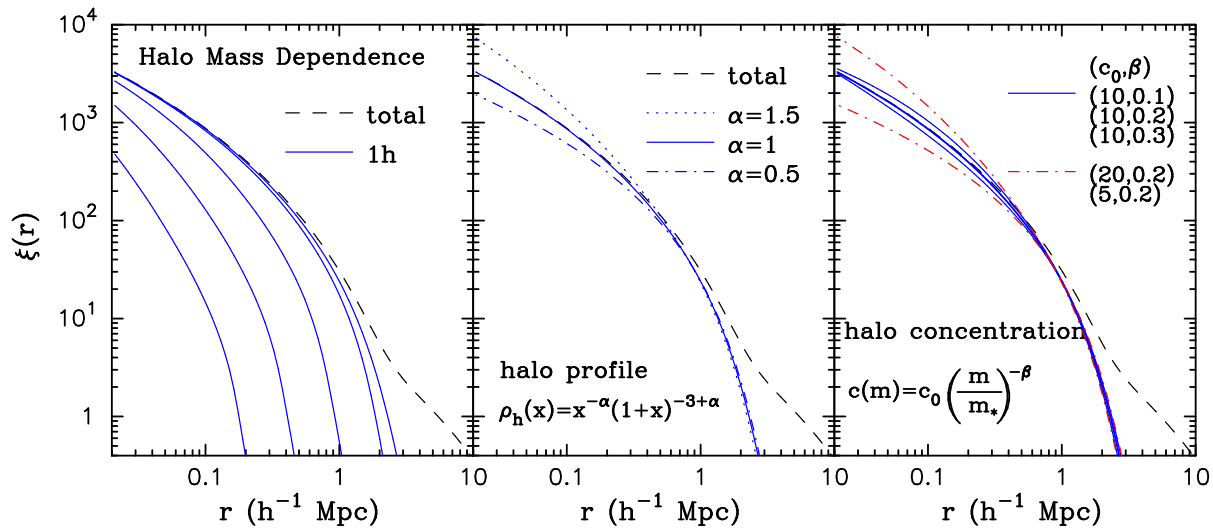}
  \end{center}
\caption{(a) The left panel shows the dependence of the 1-halo 
contribution to $\xi(r)$ on the maximum mass cutoff used in the calculation. 
From top to bottom, the maximum mass is $10^{16}$, 
$10^{15}$, $10^{14}$, $10^{13}$ and $10^{12}M_\odot$. For comparison, 
the dashed curve denotes the full $\xi(r)$ (1h+2h) for our reference
 model. 
(b) In the middle panel, the three solid curves demonstrate
 the dependence on the inner slope parameter $\alpha$ of the 
 halo profile (see equation
 (\ref{eqn:nfw})). 
(c) The right panel 
shows the dependence on the concentration parameter. 
The result for our  fiducial model $(c_0,\beta)=(10,0.2)$ is shown by the 
thick solid curve.      
} \label{fig:2ptdep}
\end{figure}
The left panel of Figure \ref{fig:2ptdep} shows the dependence
on the maximum mass cutoff used in the calculation on $\xi(r)$. From top
to bottom, the solid curves show the results for the maximum cutoff of
$10^{16}$, $10^{15}$, $10^{14}$, $10^{13}$ and $10^{12}M_\odot$. Massive
halos with $m\simgt 10^{14}M_\odot$ yield the dominant contribution
(greater than $\sim 80\%$) to $\xi_{1h}(r)$ for $r\simgt
0.1~h^{-1}{\rm Mpc}$, while less massive halos become relevant at 
smaller scales. The dashed curve denotes the full halo model prediction 
for $\xi(r)$ (1h+2h) as shown in Figure \ref{fig:comp_2pt}.

The middle and right panels of Figure \ref{fig:2ptdep} 
show the dependences of $\xi(r)$ on the
variations in the halo profile. From the middle panel, one can see that
increasing $\alpha$ for the halo profile of equation (\ref{eqn:nfw}) leads to a
higher amplitude and steeper slope for $\xi(r)$ on small scales
(see also Jain \& Sheth 1997; Ma \& Fry 2000c). For generic $\alpha$,
the angular integration in $\int d^3\bm{s}$ in equation (\ref{eqn:r2ptfull})
is not analytic, so we performed the 3-dimensional integration to get
$\xi_{1h}(r)$ numerically.  The right panel shows the dependences
on the concentration parameter parameterized in terms of the
normalization $c_0$ and the slope $\beta$ as
$c(m)=c_0(m/m_\ast)^{-\beta}$.  The result of our
fiducial model $(c_0,\beta)=(10,0.2)$ is shown by the thick solid curve.
It is apparent that increasing $c_0$ with fixed $\beta$ or decreasing
$\beta$ with fixed $c_0$ increases $\xi$ at small
scales. This is because these variations lead to more concentrated
density profiles for halos more massive than the nonlinear mass scale
$m_\ast$.  Since the massive halos dominate the contribution to $\xi$ at
the scales considered here, this has the effect of increasing $\xi$. The
results in Figure \ref{fig:2ptdep} imply that we can adjust the inner
slope $\alpha$ and the concentration parameter, both of which are
somewhat uncertain theoretically and observationally, so that the halo
model can reproduce the PD result for a given mass function (see Seljak
2000 for a similar discussion).  We will argue that the 3PCF can be used
in combination with the 2PCF to constrain properties of the halo
profile.
 
\begin{figure}
  \begin{center}
    \leavevmode\epsfxsize=8.4cm \epsfbox{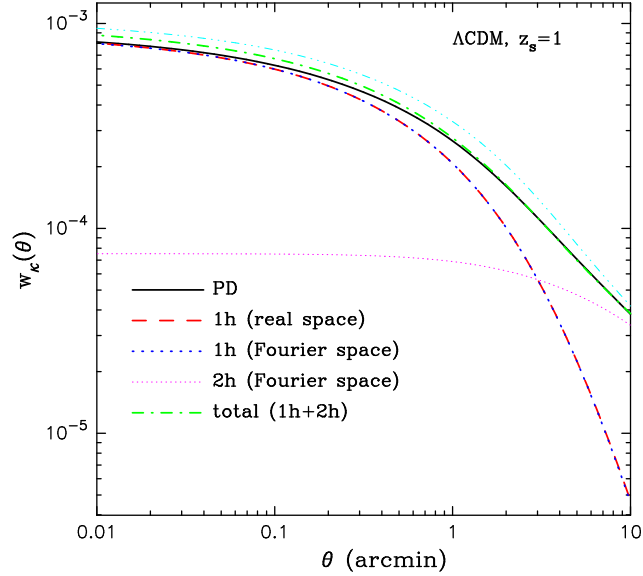}
  \end{center}
\caption{Comparison between model predictions of the two-point
correlation function of the weak lensing convergence field, as in Figure
\ref{fig:comp_2pt}. The dashed curve is the real-space halo model
prediction (equation \ref{eqn:2ptconv}) for the 1-halo contribution to the 
2PCF, while the thick dotted curve denotes the result of the Fourier-space halo
model computed from equation (\ref{eqn:2d2pt}). The top, broken curve shows
the overestimation of the halo model result (1h+2h), if the NFW
convergence field in our equation (\ref{eqn:2ptconv}) is replaced
by Bartelmann's (1996) result.  } \label{fig:ang2pt}
\end{figure}
Figure \ref{fig:ang2pt} shows the 2PCF
of the weak lensing convergence field. The real-space halo
model prediction (dashed curve) for the 1-halo contribution is compared
with the result of the
Fourier-space halo model (thick dotted curve), 
computed from equations (\ref{eqn:2ptconv}) and (\ref{eqn:2d2pt}),
respectively. Here we have considered the NFW profile and source
redshift $z_s=1$ for simplicity. The different halo model
predictions perfectly agree over the
angular scales we have considered. This agreement is encouraging,
because the real-space halo approach can then be used to predict the
3PCF with the same accuracy expected as for the Fourier-space model well
studied in the literature (TJ02 and references therein).  For
comparison, the solid curve denotes the PD result and the thin dotted
curve is the Fourier-space halo model prediction of the 2-halo term.
It clarifies that the 2-halo term is important on $\theta\simgt 3'$ for
the \LCDM model.  The dot-dashed curve is the total contribution to the
2PCF from the 1- and 2-halo terms, which slightly deviates from the PD
result on the small scales $\theta\simlt 1'$. As discussed below
equation (\ref{eqn:formsig}), if we use the expression in Bartelmann
(1996) for the NFW convergence field in the calculation of the
1-halo term, one finds discrepancies between the real-space and
Fourier-space halo models. The broken curve shows the result (1-halo +
2-halo terms), which overestimates $w_{\kappa}$ by 10-25 \% over the
scales considered. 

\subsection{3-point correlation function}

\subsubsection{Dark matter correlation function}
\begin{figure}
  \begin{center}
    \leavevmode\epsfxsize=6.cm \epsfbox{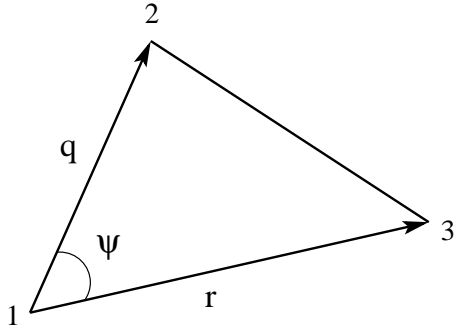}
  \end{center}
\caption{The sketch of the triangle configuration parameterized 
by $r$, $q$ and $\psi$ used in this paper to describe
the 3PCF.
} \label{fig:triang}
\end{figure}
We first consider the 3PCF of the mass
density field.  To obtain the 1-halo contribution,
where all the three matter particles reside in one halo, we perform
the following 4-dimensional numerical integration;
\begin{eqnarray}
\zeta_{1h}(r,q,\psi)=
\int\!\!dm~ n(m)\left(\frac{m}{\bar{\rho}_0}\right)^3
\int_0^{\rvir}\!\!ds\int^{2\pi}_0\!\!d\varphi
\int^\pi_0\!\!d\theta~ s^2\sin\theta~
 u_m(s)u_m(|\bm{s}+\bm{r}|)u_m(|\bm{s}+\bm{q}|),
\label{eqn:r3ptcalc}
\end{eqnarray}
where $|\bm{s}+\bm{r}|=(s^2+r^2+2sr\cos\theta)^{1/2}$ and
$|\bm{s}+\bm{q}|=[s^2+q^2+2sq(\sin\psi\sin\theta\cos\varphi
+\cos\psi\cos\theta)]^{1/2}$. Note that we employ the boundary condition
$u_m(r)=0$ for $r> \rvir$, but this does not affect the final results
on the non-linear scales, as explained in Figure \ref{fig:comp_2pt} and
explicitly demonstrated in Figure \ref{fig:excl}.  In the
above equation, we have used the fact that the 3PCF can be expressed
as a function of the three independent parameters $r$, $q$ and $\psi$
specifying the triangle configuration (see Figure \ref{fig:triang}).

To complete our halo model predictions, we develop 
approximations for calculating the 2- and 3-halo terms in Appendix
\ref{app3pt}.  The 2-halo term is
relevant only for small range of scales in the transition 
between the quasi-linear and
non-linear regimes ($r\sim 1~ h^{-1}{\rm Mpc}$), while the 3-halo term is
dominant on larger scales. 
As in the literature, we consider the hierarchical 3PCF amplitude
defined as
\begin{equation}
Q(r,q,\psi)=\frac{\zeta(r,q,\psi)}
{\xi(r)\xi(q)+\xi(r)\xi(|\bm{r}-\bm{q}|)+\xi(q)\xi(|\bm{r}-\bm{q}|)},
\label{eqn:qpara}
\end{equation}
where $\zeta=\zeta_{1h}+\zeta_{2h}+\zeta_{3h}$. 
In the following halo model predictions for $Q$, we will maintain 
self-consistency by using
the halo model for calculating the 2PCF in the denominator of $Q$.

\begin{figure}
  \begin{center}
    \leavevmode\epsfxsize=14.cm \epsfbox{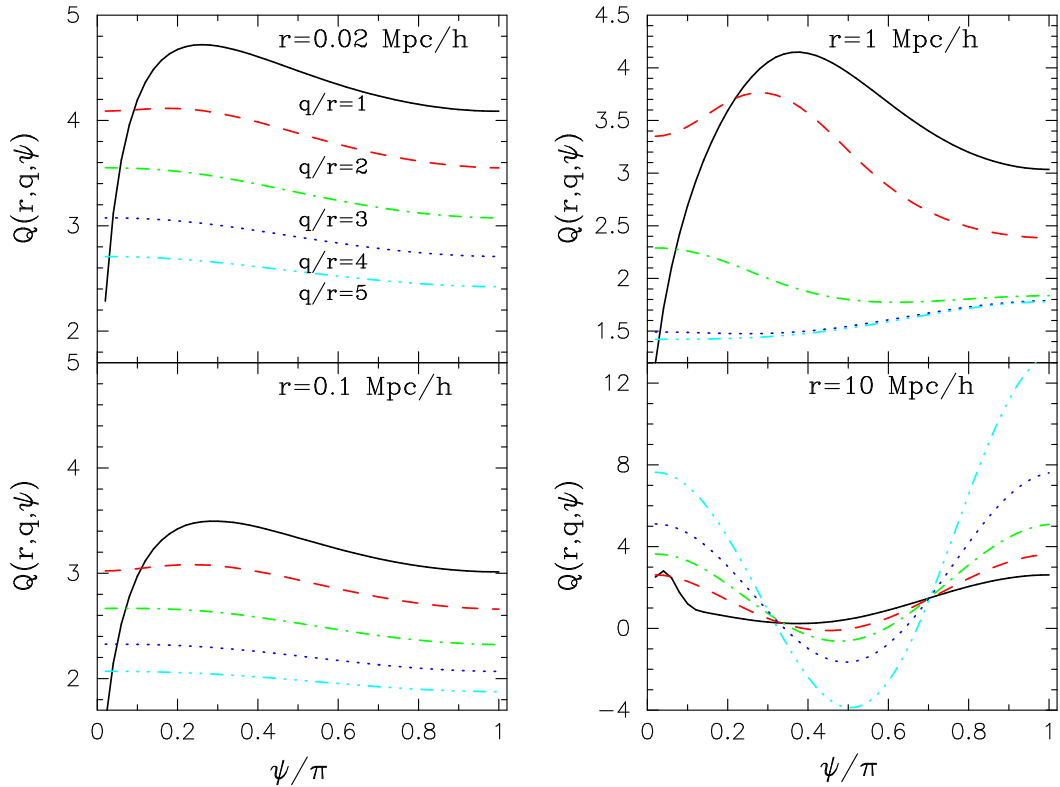}
  \end{center}
\caption{The halo model predictions for the reduced 3PCF, $Q$, as a
function of the shapes of triangle configurations. The results for
the NFW profile at $z=0$ are plotted for the \LCDM model.  The
triangle shape is parameterized by three parameters $r$, $q$ and $\psi$
as illustrated in Figure \ref{fig:triang}.  The four panels show the
results for $Q$ vs. $\psi$, with $r=0.02$, $0.1$, $1.0$ and 
$10~ h^{-1}$ Mpc as indicated in the panels. In each
panel, the solid, dashed, dot-dashed, dotted, and broken curves denote
the $Q$ parameter for $r/q=1$, $2$, $3$, $4$ and $5$, respectively. } 
\label{fig:qpara}
\end{figure}
Figure \ref{fig:qpara} plots the dependences of $Q(r,q,\psi)$ on the
size and shape of triangle configurations. In this figure we have used the NFW
profile.  The four panels show the results for $r=0.02$, $0.1$, $1$ and
$10~ h^{-1}{\rm Mpc}$. In each panel, the solid, dashed, dot-dashed,
dotted and broken curves plot the $Q$ values for $q/r=1$, $2$, $3$, $4$
and $5$ as a function of the interior angle $\psi$. 
It is apparent that on the highly non-linear scales
$r\simlt 0.1~ h^{-1}{\rm Mpc}$ the 3PCF has a weak dependence on the
triangle configuration; even so, for plausible halo model parameters 
the hierarchical form $Q\approx {\rm
const.}$ does not hold in that $Q$ depends both on the size and shape
of the triangle configuration (also see discussion in Ma \& Fry 
2000b). The curves with $q/r=1$ show
that $Q$ decreases with decreasing $\psi$ at $\psi \simlt
0.2\pi$ because of the strong suppression due to $\xi^2$ in the
denominator of $Q$. On larger scales $r\simgt 1~{\rm Mpc}$, an
oscillatory feature in $Q$ appears as predicted by 
perturbation theory (e.g., Fry 1984b; Barriga \& Gazta\~naga 2002).
Thus, the configuration dependence of the 3PCF is more prominent on
quasi-linear scales than on strongly non-linear scales.
 
\begin{figure}
  \begin{center}
    \leavevmode\epsfxsize=14.cm \epsfbox{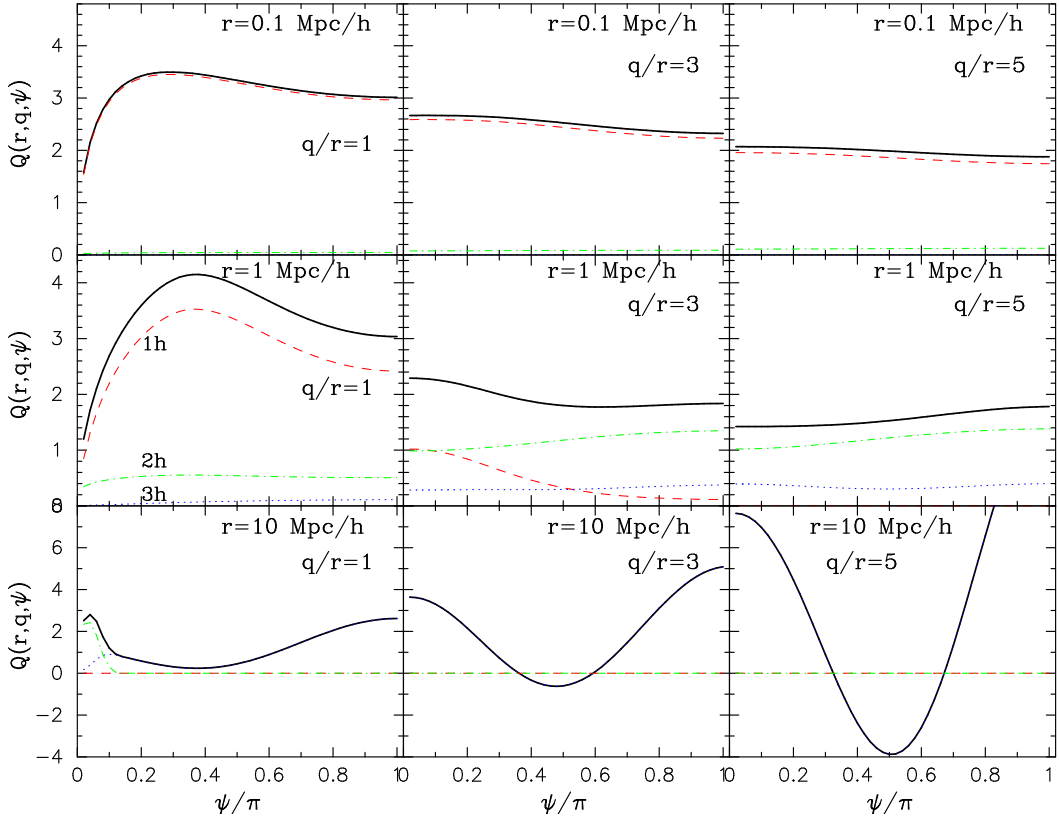}
  \end{center}
\caption{The 1-, 2- and 3-halo contributions to the
$Q$ parameter are shown by the dashed, dot-dashed and dotted curves, 
respectively. The triangle shapes are parameterized 
as in Figure \ref{fig:qpara}. Note that in the top panel the 1-halo
term dominates $Q$, while in the bottom panel the 3-halo term dominates. 
} \label{fig:qpara1h}
\end{figure}
The features observed in $Q$ can be understood using Figure
\ref{fig:qpara1h}, which shows the 1-, 2- and 3-halo contributions to
$Q$ separately for the triangle configurations in Figure
\ref{fig:qpara}.  Note that these separate contributions are shown
for the 3PCF, the numerator of $Q$, while the 2PCF factors in the
denominator include the full contribution from the 1- plus 2-halo terms.
This figure clarifies that the 1-halo term yields the dominant contribution
to the 3PCF on small scales $r\simlt
1~{\rm Mpc}$.  The 2-halo term becomes relevant over the
transition scales $1\simlt r\simlt 5~ {\rm Mpc}$ between the non-linear 
and linear regimes, while the 3-halo term dominates for $r\simgt
10~ {\rm Mpc}$.

\begin{figure}
  \begin{center}
    \leavevmode\epsfxsize=14.cm \epsfbox{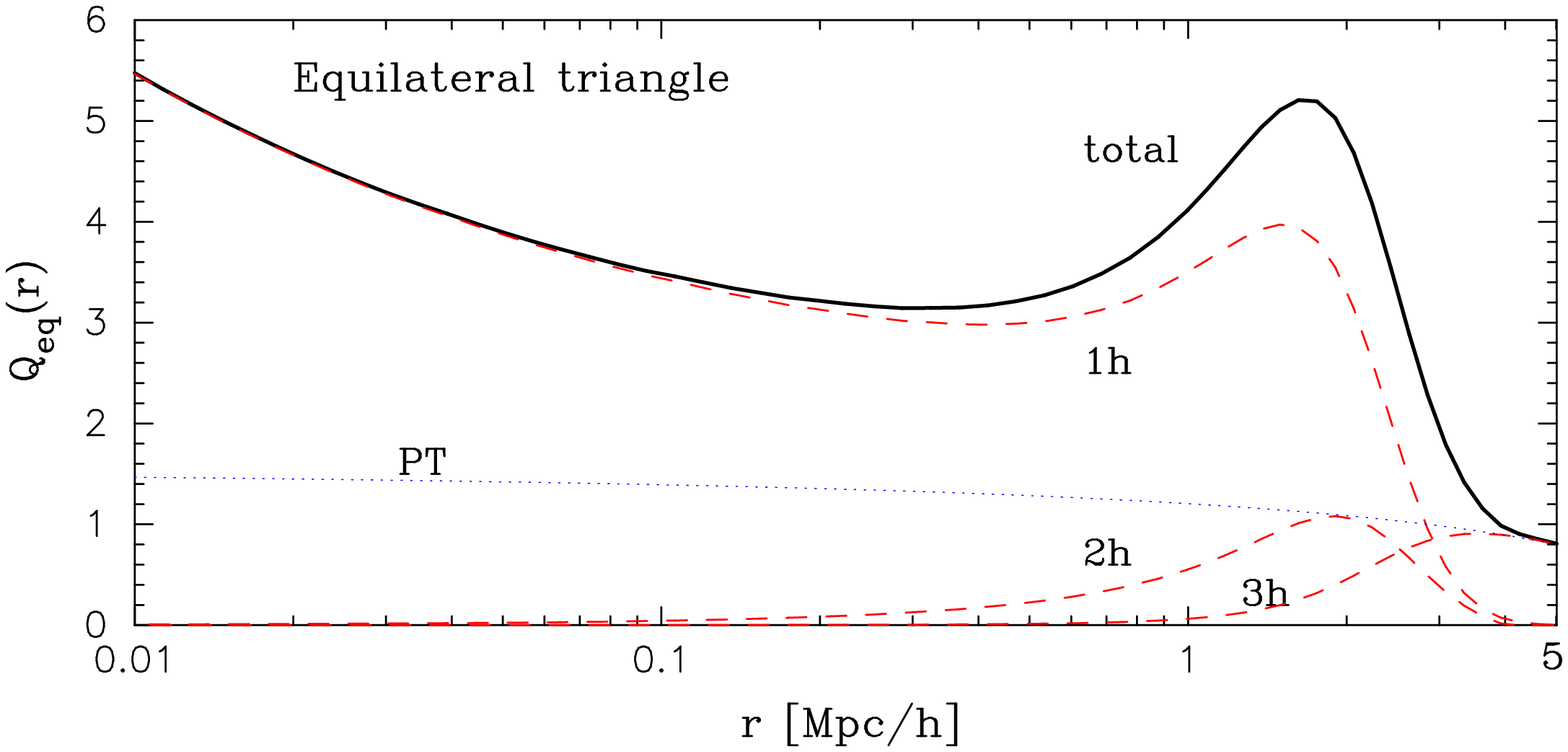}
  \end{center}
\caption{  Halo model predictions for the reduced 3PCF for equilateral
triangles as a function of the side length $r$. 
The three dashed curves represent the 1-, 2- and 3-halo contributions
 separately.  For
 comparison, the dotted curve is the perturbation theory result, which
 matches the 3-halo term on large scales  $\simgt 3~ h^{-1}$Mpc. 
} 
\label{fig:qparauv}
\end{figure}
Figure \ref{fig:qparauv} displays the halo model predictions for $Q$ for
equilateral triangles as a function of the side length $r$.  The three
dashed curves plot the 1-, 2- and 3-halo contributions separately.  For
comparison, the perturbation theory result is shown by the dotted curve.
The figure again clarifies that, at the small scales $r\simlt
0.5~h^{-1}{\rm Mpc}$, the 1-halo term dominates the total contribution,
while the 3-halo term captures the large-scale correlations in the
quasi-linear regime. A comparison between the perturbation theory result
and the 3-halo term reveals that our approximation (\ref{eqn:app3h})
reproduces the perturbation theory result at large scales $r\simgt 5~
h^{-1}{\rm Mpc}$. Thus the way the halo model provides a separate
description of the 1-, 2- and 3-halo terms can clarify how gravitational
clustering transits from the linear regime to the strongly non-linear
regime as one goes to smaller scales.

The halo model predictions match $N$-body
simulation results on small scales $r\simlt 1~
h^{-1}{\rm Mpc}$ (R. Scoccimarro; private communication).  However, 
the  $Q$ parameter from the halo model has a bump
feature at $r\simeq 2~h^{-1}{\rm Mpc}$ which does not seem to be 
present in simulation data.  
The corresponding bump
feature in the bispectrum has been found at $k\sim 1~ h{\rm
Mpc}^{-1}$ in previous work (see Figure 2 in Scoccimarro et
al. 2001 and Figure 13a in Cooray \& Sheth 2002). 
Hence, it is unlikely to be due to inaccuracies in our
approximations used for the 2- and 3-halo term calculations, 
although the approximations tend to overestimate the true amplitudes 
to some extent as carefully investigated in TJ02. 
It probably does not reflect real properties of dark
matter clustering either (as discussed below for Figure
\ref{fig:qjing}).  Rather, it can be ascribed to an inaccuracy 
in the standard implementation of the halo model which appears 
only on transition scales of order a few Mpc, in
contrast to the success of the halo model in the
non-linear and linear regimes.

\begin{figure}
  \begin{center}
    \leavevmode\epsfxsize=14.cm \epsfbox{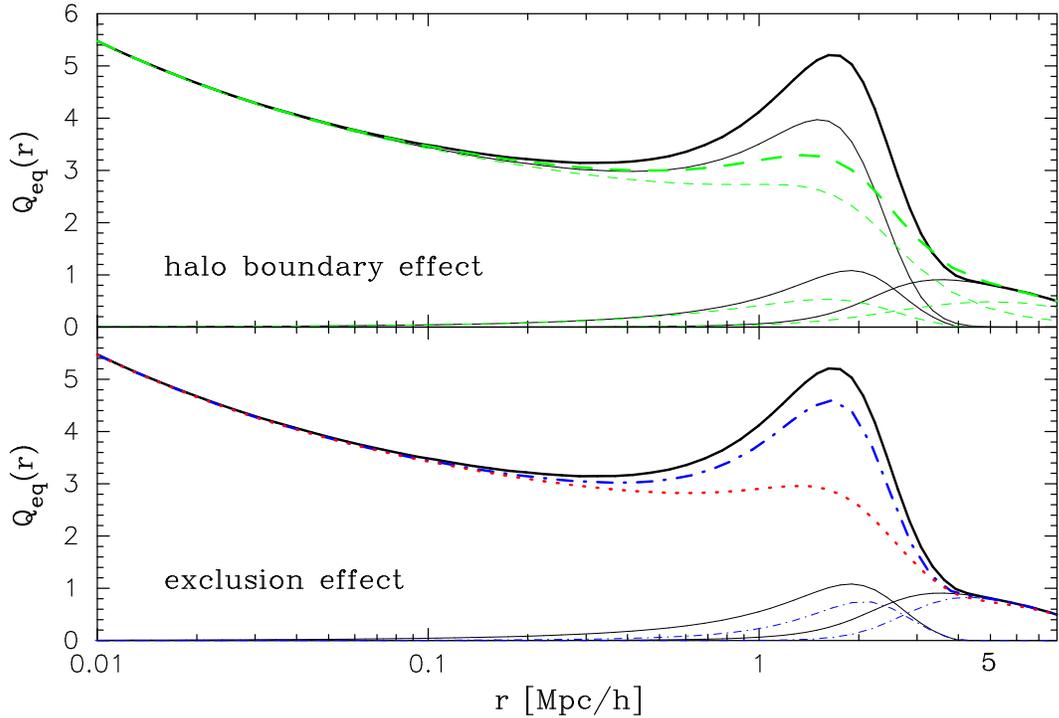}
  \end{center}
\caption{This figure shows the effect of modifying the halo model
calculation by altering the halo boundary condition and by exclusion
effects in the 2- and 3-halo terms. 
The upper panel shows how the bump in
$Q_{\rm eq}(r)$ is suppressed
if halos are not truncated at the virial radius (see text for details). 
The thick dashed curve shows $Q_{\rm eq}(r)$
and the three thin dashed curves show the 1-, 2- and 3-halo contributions.
For comparison, the solid curves are the halo model results shown in 
Figure \ref{fig:qparauv}, in which a sharp cutoff is used 
at the virial radius. 
The dot-dashed curve in the lower panel shows 
the effect on $Q_{\rm eq}(r)$ if halo exclusion
effects are included in calculations.
The 2-halo term
 of the 2PCF and the 2- and 3-halo terms of the 3PCF, which enter 
denominator and numerator in $Q_{\rm eq}(r)$, are constrained by 
the condition  $r_{\rm vir}(m)\le r/2$. 
This effect suppresses 
the amplitudes of the 2- and 3-halo contributions to $Q_{\rm eq}$, again leading to 
the suppression of the bump feature. The 1-halo term is not shown
in this panel as it remains unaffected. The thick dotted curve shows
the prediction for $Q_{\rm eq}(r)$ if both boundary and exclusion effects
are included. 
} \label{fig:excl}
\end{figure}
One possible origin of the inaccuracy is the sharp cutoff at the virial
radius for the integrals for the 2PCF and 3PCF,
as have been used for the real-space halo model 
in this paper as well as for the Fourier-space model 
in Scoccimarro et al (2001) and Cooray \& Sheth (2002). 
As shown in Figure
\ref{fig:comp_2pt}, modifications of the integration range could
alter the halo model prediction at the transition scales.  
In the upper panel of Figure \ref{fig:excl}, the thick 
dashed curve shows the result 
obtained without using the boundary
condition  $u_m(x)=0$ for $x>\rvir$ for the calculations of the 1-halo
terms of the 2PCF and 3PCF, which enter the denominator and numerator of
$Q$, respectively.  One can see that the bump feature is weakened
through complex dependences -- the 2PCF amplitude is enhanced by the
modified boundary condition more strongly than the  3PCF.
In fact, this trend is verified by the results in Figure 3 and 4 in Ma
\& Fry (2000b), which show that the Fourier-space halo model yields 
bispectra with {\em no} bump feature.  Ma \& Fry employed 
the Fourier transform of the
non-truncated profile, $\tilde{u}_m(k)$, calculated over an  infinite
integration range, which includes mass contribution outside
the virial region.  However, in this paper we
have used the virial boundary condition since it preserves
mass conservation.
We have also clarified in Figure \ref{fig:ang2pt} that, if we include
mass contributions beyond the virial radius, the halo model
prediction for the lensing 2PCF overestimates the 
amplitude (see the dot-dashed and broken curves in Figure
\ref{fig:ang2pt}).  In reality, the mass distribution outside the virial
region eventually merges into quasi-linear structures, such as filamentary
structures in the universe, which are relevant for the transition scale
($\delta\simgt 1$), and are unlikely to follow the halo profile
far outside the virial region.
Therefore, a careful investigation of which boundary conditions to use
for the halo model integrals will be needed to achieve more accurate
predictions at the transition scale.

Another effect that could be manifested at these scales is 
the exclusion effect of different halos 
for the 2- and 3-halo terms. We could impose the condition that
different halos  are 
separated by scales larger than the sum of their 
virial radii. The halo model does not explicitly
account for this effect.  Since the transition scale $\sim 1\ {\rm Mpc}$ is
of the order of the virial radius of massive halos that make a significant
contribution to the integrals, the halo model used
here might  overestimate the contribution from the 2- and
3-halo terms at these scales.  
The dot-dashed curve
in the lower panel of Figure \ref{fig:excl} shows an estimate of this 
effect. The estimate is obtained by imposing the condition 
$r_{\rm vir}(m)\le r/2$, an approximate prescription to exclude
pairs and triplets of halos contributing to the 2- and 3-halo terms at scales
smaller than the sum of their virial radii. 
The plot shows that including the exclusion effect suppresses the
bump feature in $Q$. The dotted curve in the lower panel of
Figure  \ref{fig:excl} shows the result of applying both alterations
(boundary and exclusion effects) to the calculations. 
A systematic resolution of  these problems is
beyond the scope of our paper, and will be considered elsewhere. 
Finally, the figure implies that, except for the problematic
bump feature, a  transition between the non-linear
amplitude ($Q\simgt 3$) and the quasi-linear amplitude ($Q\simlt 1$) occurs
at scales of a few $ h^{-1 }{\rm Mpc}$, corresponding to a flattening
of the 2PCF $\xi(r)$ (see Figure \ref{fig:comp_2pt}).

\begin{figure}
  \begin{center}
    \leavevmode\epsfxsize=14.cm \epsfbox{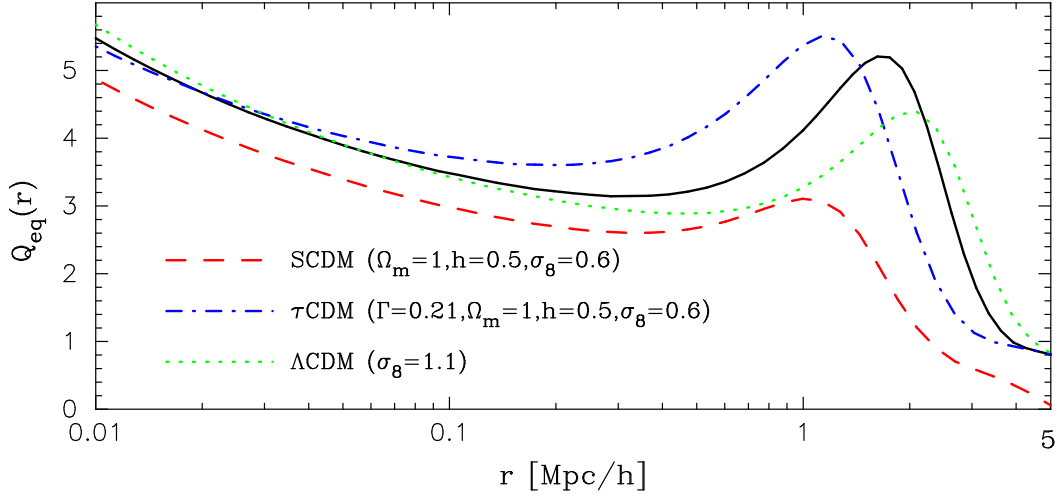}
  \end{center}
\caption{
The results for $Q_{\rm eq}(r)$ are shown 
for different cosmological models as in Figure \ref{fig:qparauv}. 
The dotted curve denotes the \LCDM model with 
$\sigma_8=1.1$, compared to $\sigma_8=0.9$ for our
fiducial model. The dashed curve shows the
result for the SCDM model. To show the sensitivity of $Q$ to the shape of the
 matter power spectrum, we also show the result for the $\tau$CDM model,
 which is different only in the shape parameter ($\Gamma=0.21$)
from the SCDM model ($\Gamma=0.5$).
} \label{fig:qcosmo}
\end{figure}
Figure \ref{fig:qcosmo} shows the dependences
of $Q$ on the cosmological model. We consider three models that differ from
our fiducial model.  One is the $\Lambda$ CDM model normalized with
$\sigma_8=1.1$, motivated by the possible detection of the
Sunyaev-Zel'dovich effect in the CMB (Bond et al. 2002). The second is
a flat model without cosmological constant (SCDM), with 
parameters  $\Omega_{m0}=1$, $h=0.5$ and $\sigma_8=0.6$. 
To clarify the dependence on the shape of the matter power spectrum, 
we also consider the $\tau$CDM model, where the shape parameter
$\Gamma=0.21$, but the other parameters are same as in the SCDM
model.   The figure reveals that the $Q$ parameter is not very different
for the different models but does vary slightly with changes in the various
model parameters. Thus accurate measurements must be interpreted with
precise predictions that explore all parameter dependences. 

\begin{figure}
  \begin{center}
    \leavevmode\epsfxsize=14.cm \epsfbox{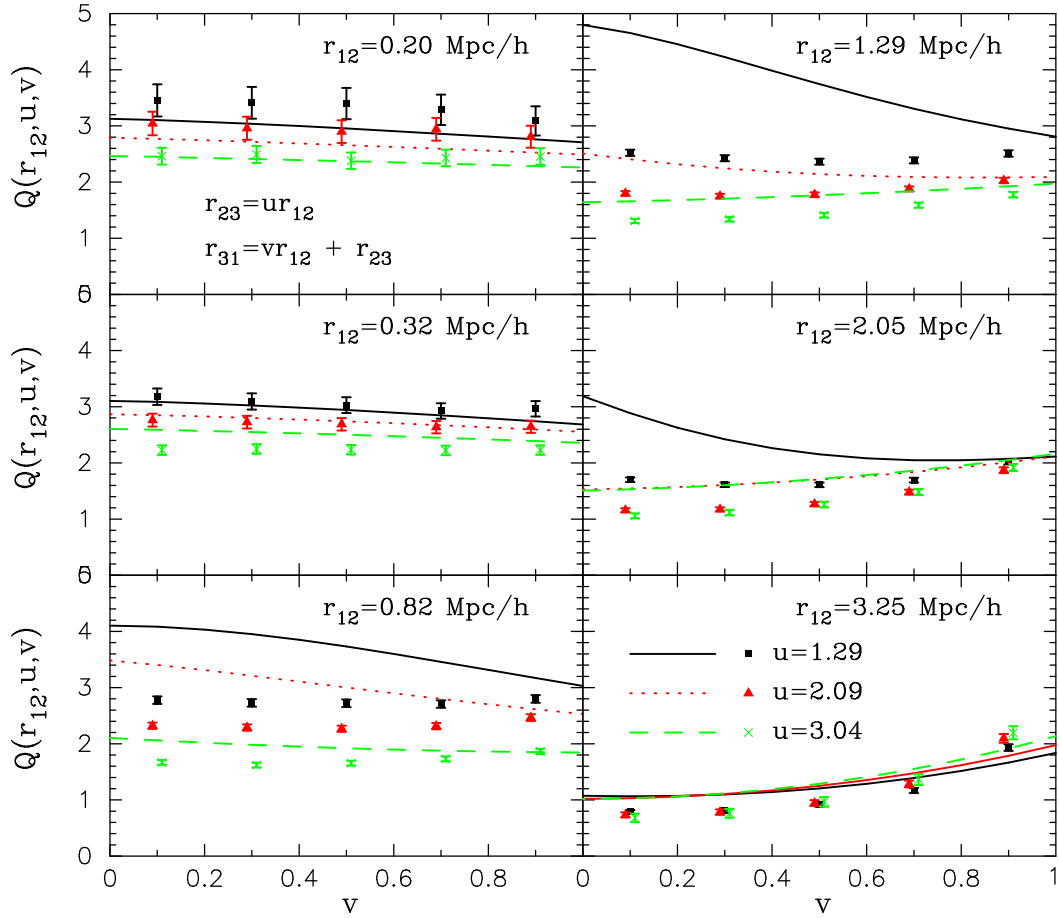}
  \end{center}
\caption{Halo model predictions for the dependence of the $Q$
parameter on  the triangle shapes parameterized by $r_{12}$, $u$ and
 $v$ (see equation (\ref{eqn:uv})).   The values of $r_{12}$, $u$ and $v$
are the same as those used in Figure 2 in Jing \& B\"orner (1998), where
the $Q$ parameters were computed from $N$-body simulations. 
The data with error bars denote the simulation results. 
} \label{fig:qjing}
\end{figure}
For completeness we present our results with the triangles specified
using a set of alternative parameters which have been 
used in the literature (e.g., Peebles 1980):
\begin{equation}
u\equiv \frac{r_{23}}{r_{12}},\hspace{2em} v\equiv\frac{r_{31}-r_{23}}{r_{12}},
\label{eqn:uv}
\end{equation}
with the condition $r_{12}\le r_{23}\le r_{31}$ which imposes the
constraints $u\ge 1$ and $0\le v\le 1$. The 3PCF for any triangle
can then be expressed as a function of three parameters $r_{12}$, $u$ and $v$,
where $u$ and $v$ characterize the shape and $r_{12}$ the size of a
triangle.

Figure \ref{fig:qjing} shows the dependence of $Q$ on 
$r_{12}$, $u$ and $v$. The values of $r_{12}$,
$u$ and $v$ are chosen such that our model predictions can be compared
with the $N$-body simulation results shown in Figure 2 in Jing \&
B\"orner (1998).  
The data with the error bars represent the simulation
results, which were kindly made available to us by Yipeng Jing. 
Although the cosmological parameters in Jing \& B\"orner (1998)
are slightly different from our model, this fact does not matter because
of the weak dependence of $Q$ on the cosmological model as explained in
Figure \ref{fig:qcosmo}.  
It is also noted that the halo model prediction does not account for the
effect of averaging over bin widths in $u$ and $r_{12}$ which is taken
for the $Q$ measurement from the simulations. 
This figure reveals that our halo model
agrees well with the simulation results for the configuration
dependence of $Q$ and its amplitude at the non-linear scales 
$r_{12}=0.2$ and $0.32~h^{-1}$Mpc, as well as the quasi-linear scale 
$r_{12}=3.25~h^{-1}$Mpc.  However, as discussed above,
for the case in which one side length of the triangle is
comparable to $1~ h^{-1}{\rm Mpc}$, the halo model tends to overestimate
the simulation result.  More detailed comparisons with recent higher
resolution simulations will be presented elsewhere. 

\begin{figure}
  \begin{center}
    \leavevmode\epsfxsize=8.cm \epsfbox{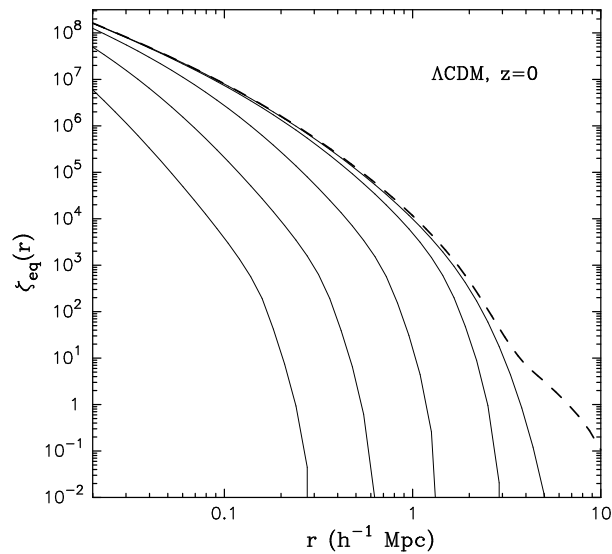}
  \end{center}
\caption{The dependence of the mass 3PCF on the maximum mass
 cutoff used in the halo model calculations, plotted against the side
 length of the equilateral triangle. From top to bottom, 
the maximum mass is $10^{16}$, 
$10^{15}$, $10^{14}$, $10^{13}$ and $10^{12}M_\odot$.
The dashed curve is the full 3PCF.  
} \label{fig:3ptmmax}
\end{figure}
Figure \ref{fig:3ptmmax} shows how the 1-halo contribution to the mass
3PCF depends on the maximum mass cutoff used in the calculation against
the side length of equilateral triangles. This can be compared to the
dependence of the 2PCF on mass cutoff shown in the left panel of Figure
\ref{fig:2ptdep}. For the 3PCF more massive halos dominate the
contribution at a given length scale: for example, at $r=0.5~h^{-1}$Mpc, 
over half the contribution to the 3PCF 
is from halos with $m>10^{15}M_\odot$, whereas for the 2PCF the mass
range is $m>10^{14}M_\odot$. 

\begin{figure}
  \begin{center}
    \leavevmode\epsfxsize=12.cm \epsfbox{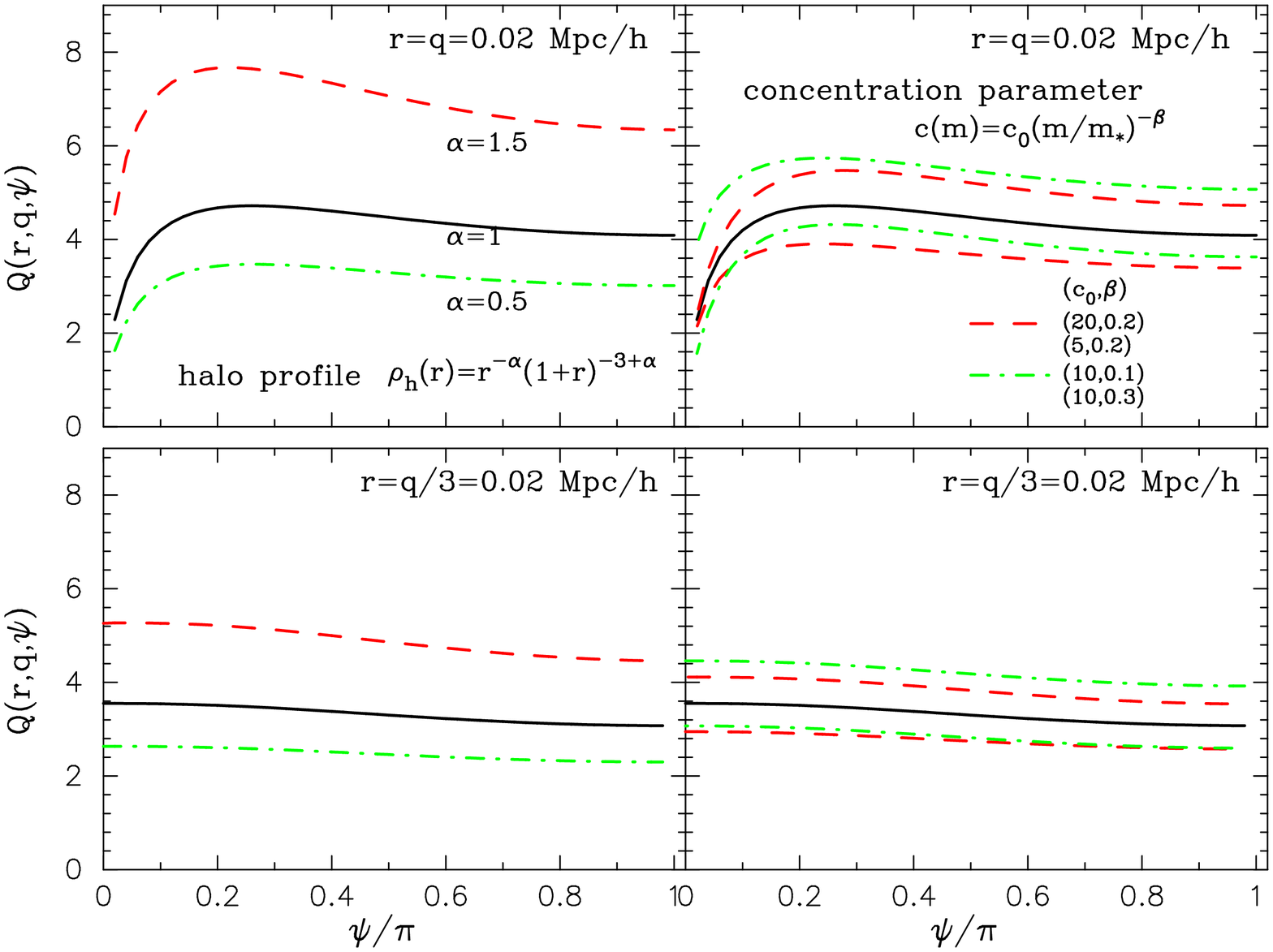}
  \end{center}
\caption{The dependences of the $Q$ parameter on the halo profile
parameters: the 
inner slope $\alpha$ and the concentration parameter, $c(m)$,
parameterized in terms of the normalization $c_0$ at the nonlinear mass
scale, and the slope $\beta$. The upper panels show the results for
$Q$ versus $\psi$ with the two side lengths
$r=q=0.02~ h^{-1}{\rm Mpc}$ 
(Figure \ref{fig:triang} defines $\psi, r, q$). 
The lower panels have
$r=q/3=0.02~ h^{-1}{\rm Mpc}$.  
} 
\label{fig:qprofr001}
\end{figure}
\begin{figure}
  \begin{center}
    \leavevmode\epsfxsize=12.cm \epsfbox{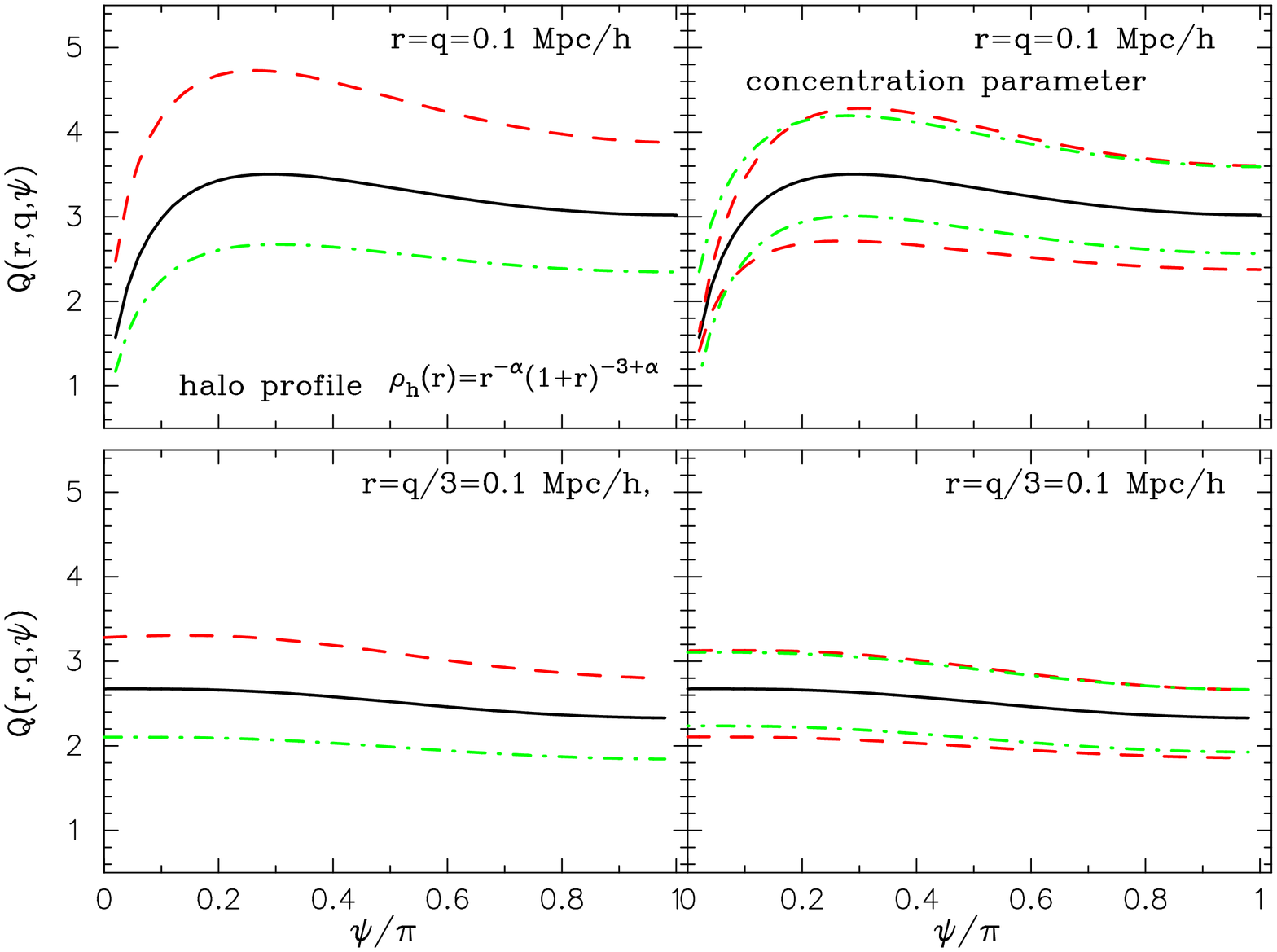}
  \end{center}
\caption{
The $Q$ values, as in the previous figure, for $r=0.1~ h^{-1}{\rm Mpc}$.} 
\label{fig:qprofr01}
\end{figure}
\begin{figure}
  \begin{center}
    \leavevmode\epsfxsize=10.cm \epsfbox{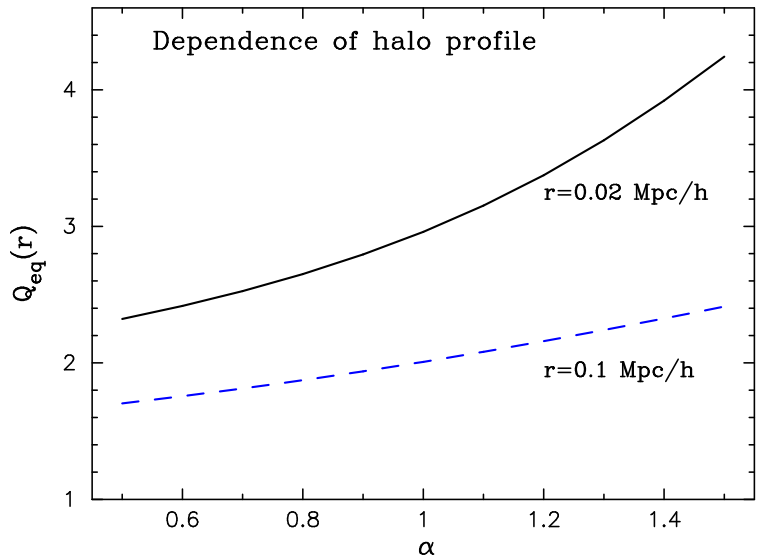}
  \end{center}
\caption{The $Q$ parameter for equilateral triangles as a
function of the inner slope parameter $\alpha$ for the halo density
 profile given by equation
(\ref{eqn:nfw}). The solid and dashed curves are the results for
 $r=0.02$ and $0.1~ h^{-1}$Mpc, respectively.}
\label{fig:qalpha}
\end{figure}

In Figure \ref{fig:qprofr001} and \ref{fig:qprofr01}, we present
dependences of the $Q$ parameter on possible variations in the halo
inner slope and the concentration parameter, as in the middle and right
panels of Figure \ref{fig:2ptdep}.  To do this, we use the
parameters $r$, $q$ and $\psi$ for the triangle configuration as in
Figure \ref{fig:qpara}. Note that the results shown are computed from
only the 1-halo contributions to the 3PCF and 2PCF, which enter the
numerator and denominator of $Q$, respectively, since the variations of
the halo profile are relevant only at the non-linear scales.  
Increasing the inner slope parameter $\alpha$ for the profile
(\ref{eqn:nfw}) leads to higher $Q$
(see similar discussions in Ma \& Fry 2000a for the bispectrum). 
A comparison of
Figures \ref{fig:qprofr001} and \ref{fig:qprofr01}
shows that $Q$ is more sensitive to $\alpha$ 
for triangles with smaller size.  Compared with the effect on the 2PCF
shown in Figure \ref{fig:2ptdep}, this result also means that the 3PCF is
more sensitive to the inner slope of halo profile than the 2PCF.  We
find the following fitting formula of the dependence of the 
2PCF and the 3PCF on $\alpha$, applicable over the 
range $0.5\le \alpha\le 1.5$;
\begin{eqnarray}
\xi^2(r)=[\xi_{NFW}^2]^{1+0.017(\alpha-1)+0.074(\alpha^2-1)}, \hspace{0.5em}
\zeta(r)=\zeta_{NFW}{}^{1+0.021(\alpha-1)+0.08(\alpha^2-1)},\hspace{0.5em}
Q=Q_{NFW}{}^{1+0.094(\alpha-1)+0.24(\alpha^2-1)}\hspace{2em}
\end{eqnarray}
for $r=0.02~ h^{-1}{\rm Mpc}$, while 
\begin{eqnarray}
\xi^2(r)=[\xi_{NFW}^2]^{1+0.048(\alpha-1)+0.034(\alpha^2-1)}, \hspace{0.5em}
\zeta(r)=\zeta_{NFW}{}^{1+0.06(\alpha-1)+0.033(\alpha^2-1)},\hspace{0.5em}
Q=Q_{NFW}{}^{1+0.38(\alpha-1)+0.057(\alpha^2-1)}\hspace{2em}
\end{eqnarray}
for $r=0.1~ h^{-1}{\rm Mpc}$.  Here $\xi_{NFW}$, $\zeta_{NFW}$ and
$Q_{NFW}$ denote the use of the NFW profile ($\alpha=1$);  we
have used equilateral triangle configurations for the 3PCF
calculation.  Figure \ref{fig:qalpha} explicitly plots the 
$\alpha$ dependence of $Q$.

The right panels in Figure \ref{fig:qprofr001} and \ref{fig:qprofr01}
illustrate the sensitivity of $Q$ to possible variations in the halo
concentration parameter (\ref{eqn:conc}). One can see that the change
in concentration affects the $Q$ amplitudes, but does not strongly
alter the configuration dependence.  Increasing $c_0$ with fixed $\beta$
or decreasing $\beta$ with fixed $c_0$ leads to a higher amplitude
for $Q$, which is the same trend as for the 2PCF shown in Figure
\ref{fig:2ptdep}. This implies that the 3PCF is more sensitive to the
halo concentration than the 2PCF.
These results in Figure \ref{fig:qprofr001} and \ref{fig:qprofr01} 
show that measuring the 3PCF could constrain halo profile
properties complementary to the measurements of the 2PCF,  
in analogy with
determinations of the galaxy bias parameter from measurements at
quasi-linear scales (Feldmann et al. 2001; Verde et al. 2002).

In Figure \ref{fig:qprofr001}-\ref{fig:qalpha}, we have demonstrated the
sensitivity of the mass 3PCF to the halo profile properties.  We have
 compared our halo model predictions for $Q$ with the
asymptotic shape derived in Ma \& Fry (2000b) based on the halo
model.  The asymptotic $Q$ amplitude depends only on the mass slope
$\beta$ of the concentration parameter (see equation (\ref{eqn:conc})),
the slope $\alpha'$ in the low mass end of the mass dependence of the
mass function, $n(m)\propto \nu^{\alpha'}$, and a primordial spectral
index $n$.  From equation (7) in Ma \& Fry (2000b), the asymptotic $Q$
for equilateral triangles is given by
\begin{equation}
Q(r)\propto r^{-\gamma},\hspace{1em} \gamma=\frac{\alpha'(n+3)}
{2(3\beta'+1)},
\label{eqn:ma}
\end{equation}
where $\beta'\approx 0.8\beta$ and we have considered equilateral
triangle configuration.  This equation was derived under three
assumptions: the contribution from the exponential cutoff of the mass
function in the high mass end was ignored, a scale-free primordial
power spectrum was employed and the $k$-dependence of the
Fourier-transformed halo profile, $\tilde{u}_m(k)$ was ignored.  The
third assumption leads to the consequence that the asymptotic $Q$ has no
dependence on the inner slope of the halo profile.  Since for the
profile in equation 
(\ref{eqn:nfw}) $\tilde{u}_m(k)$ does depend on the inner slope
as $\tilde{u}_m(k)\approx 1$ and $\propto k^{-3+\alpha}$ for $k\ll
c/\rvir $ and $k\gg c/\rvir$, the asymptotic $Q$ should depend on the
inner slope or more generally the halo profile shape. 
By comparing our predictions of $Q$ with the analytical result of
Ma \& Fry (2000b), we find that 
our halo model predicts a 
steeper slope for $Q(r)$ than expected from the asymptotic shape in 
equation (\ref{eqn:ma}), for plausible values of $\beta$ and $\alpha'$. 

This discrepancy is probably due to the assumptions used in analytically
deriving the asymptotic shape.  The $r$-dependence of $Q$ 
arises from a complex superposition of contributions from the
halo profile, the power spectrum shape and the shape
of the mass function. Indeed, as an example of this possibility,
Taruya, Hamana \& Kayo (2002) explicitly showed, based on the halo model,
that the non-Gaussian tail of the probability distribution function of
the density field arises from such a superposition for small smoothing
scales.

An important implication pointed out in Ma \& Fry (2000b,c) is that 
plausible halo model properties are unlikely to follow the stable
clustering hypothesis. This hypothesis is useful because it 
allows us to analytically predict the behavior
of  non-linear gravitational clustering (e.g. Peebles 1980; Jain 1997).
In fact, the popular PD fitting formula widely used in the literature
was derived based on this hypothesis (also see Hamilton et al. 1991;
Jain, Mo \& White 1995).  Very recently, Smith et al. (2002) showed that
the non-linear power spectra measured from the high-resolution $N$-body
simulations indeed showed a weak violation of the stable clustering
hypothesis. Figure 9 in Smith et al. (2002) compares the $k$-slope of the
measured non-linear power spectrum with the asymptotic prediction from
Ma \& Fry (2000b), which also shows some discrepancies.  
Hence, a careful investigation will be needed to clarify how 
non-linear gravitational clustering can be described by the 
halo model ingredients
in connection with the stable clustering hypothesis and the hierarchical
ansatz (see also the next subsection).

\subsubsection{The hierarchy of higher-order correlation functions}

As stated in \S \ref{rhalo}, the real-space halo model can be a useful
tool for predicting 
the $n$-point correlation functions in the
strongly non-linear regime, where correlations within one halo 
dominate the contribution to the $n$-point functions. 
The exact expressions for the 
the hierarchy are provided by the
BBGKY equations (Peebles 1980), 
which govern collisionless gravitational system. However, 
their complex form makes it intractable to obtain solutions 
for the higher order correlations in the non-linear regime. 
Here we will use the halo model to shed some light on a
fundamental question about gravity that has been discussed in
the literature over the last decades: 
what is the asymptotic small-scale behavior of 
the hierarchy of the $n$-point functions under gravitational 
clustering (e.g., Peebles 1980; Fry 1984a,b; Hamilton 1988; 
Ma \& Fry 2000b;
Bernardeau et al. 2002a)?

The reduced $n$-point correlation function, $Q_n$, is
defined as
\begin{equation}
Q_n\equiv \frac{
\xi^{(n)}(\bm{r}_1,\dots,\bm{r}_n)}{
\Sigma_{{\rm labelings}}\Pi_{{\rm edges}~ ij}^{n-1}\xi({r_{ij}})},
\label{eqn:qn}
\end{equation}
where the denominator is given by all  topologically 
distinct tree diagrams (the different $n^{n-2}$ ways of drawing $(n-1)$
links that connect $n$ points).  
This form is motivated by the 
expectation $\xi^{(n)}\propto \xi^{n-1}$, indicated observationally by
the pioneering measurements of the two- and three-point functions from galaxy 
surveys (e.g., Groth \& Peebles 1977; Peebles 1980), and theoretically by
perturbation theory (Fry 1984b). The amplitudes $Q_n$ are a natural 
set of statistics to describe the non-Gaussianity that results from 
gravitational clustering.

\begin{figure}
  \begin{center}
    \leavevmode\epsfxsize=10.cm \epsfbox{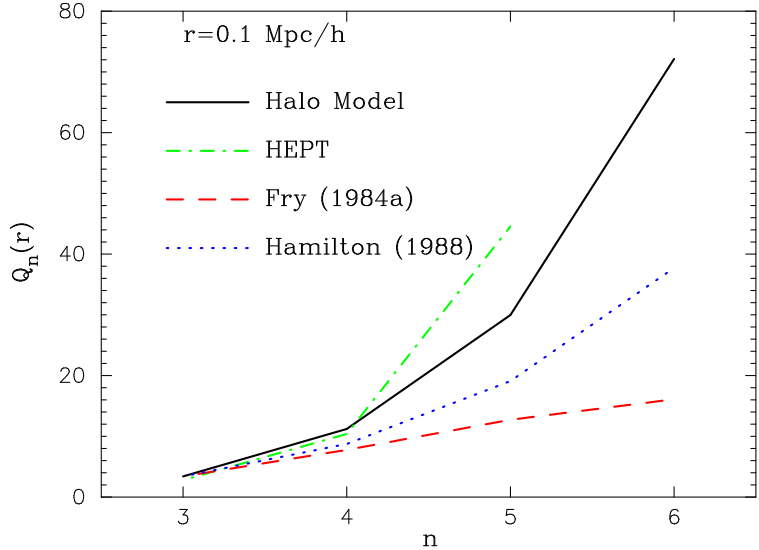}
  \end{center}
\caption{The hierarchical amplitude $Q_n$ for the mass $n$-point 
correlation function in the strongly non-linear regime is shown
versus the order $n$. 
We consider the configuration with $n$
equal sides of length $r=0.1h^{-1}$Mpc 
and equal interior angles. The solid curve shows the halo model
 prediction. The dashed curve shows the scaling for $Q_n$ as a
function of $Q_3$ and $n$ 
proposed by Fry (1984a), while the dotted curve shows the 
formula proposed in Hamilton (1988) 
(we have used the halo model result for $Q_3$ in these formulae).  
The dot-dashed curve is the prediction from hyper-extended perturbation
theory (Scoccimarro \& Frieman 1999). 
}
\label{fig:qn}
\end{figure} 
The solid curve in Figure \ref{fig:qn} 
shows the halo model predictions for 
the hierarchical amplitudes $Q_n$ for $3\le n\le6$. 
We consider configurations with $n$ equal sides, each of length $r=0.1h^{-1}$Mpc,  
and equal interior angles for evaluations of the
$n$-point functions. We take the halo profile to have the NFW form. 
For comparison, the dashed curve shows the scaling law for $Q_n$ 
proposed in Fry (1984a):  
$Q_n=(4Q_3/n)^{(n-2)}n/(2n-2)$.  The dotted curve shows the scaling 
suggested by Hamilton (1988): $Q_n=(Q_3/n)^{n-2}n!/ 2 $.    The dot-dashed
curve is the prediction from hyper-extended perturbation theory for
$3\le n\le 5$
(Scoccimarro \& Frieman 1999). 
The halo model predicts 
stronger clustering with increasing $n$ than the formulae of
Fry and of Hamilton, but not as strong as hyper-extended perturbation
theory. Note that higher-order correlations are
more sensitive to variations in the inner slope and halo concentrations 
of massive halos. 

The results in Figure \ref{fig:qn} might be compared with 
Figure 3 in Szapudi et al. (1999), which show the cumulants $S_n$ ($3\le
n \le10$) measured from the $N$-body simulations. From equation
(\ref{eqn:qn}) we can expect following rough relation between the
cumulants and the reduced $n$-point correlation functions:
\begin{equation}
S_n(r)\approx \frac{\xi^{(n)}(r)}{\left[\xi^{(2)}(r)\right]^{(n-1)}}
\sim n^{n-2}~ Q_n(r).
\label{eqn:sn}
\end{equation}
\begin{table*}
\caption{Comparison of the theoretical 
predictions for the cumulants, $S_n$, at $r=0.1h^{-1}$Mpc with the
 simulation result. 
The analytical predictions 
are estimated from the reduced $n$-point functions, $Q_n$, shown 
in Figure \ref{fig:qn}
using equation (\ref{eqn:sn}).  
The simulation results correspond to the values denoted by the dashed curves 
in Figure 3 in Szapudi et al (1999), which are 
measured from $N$-body simulations for the SCDM model. 
}
\label{tab:comp}
\begin{tabular}{lcccc}
\hline
\hline
& $S_3$ &$S_4$&$S_5$&$S_6$\\
\hline 
Halo Model & 10 & 180 & $3.7\times 10^3$ & $9.3\times 10^4$\\
HEPT (Scoccimarro \& Freeman 1999) & 8.4 & 166 & $5.6\times 10^3$ & -\\
Fry 1984a & 10 & 124 & $1.6\times 10^3$ & $2.1\times 10^4$\\
Hamilton 1988 & 10 & 139 & $2.4\times 10^3$ & $4.9\times 10^4$\\
$N$-body simulation in Szapudi et al. 1999 & $\sim 10$& $\sim 300$& 
$\sim 10^4$ & $\sim 4\times 10^5$ \\
\hline 
\hline
\end{tabular}
\end{table*}
Table \ref{tab:comp} shows a comparison of the simulation results for
$S_n$ in Szapudi et al. (1999) with the theoretical predictions scaled
from the values of $Q_n$ in Figure \ref{fig:qn}.  Note that the
simulation results were read from the dashed curves in Figure 3 in
Szapudi et al. (1999) for the SCDM model (the cosmological parameters 
are slightly different from the model we have considered), and we 
expect similar values for the $\Lambda$CDM because of the weak dependence of
$S_n$ on cosmological models.  One can see that the rapid increase
in the amplitudes of $S_n$ with $n$ seen in the simulation result is
better described by the halo model predictions than those of Fry and of
Hamilton.  However, as explicitly pointed out in Szapudi et al. (1999),
the simulation results might be overestimated due to the shot noise
effect.  Therefore, further detailed work is needed to systematically
explore the consistency of the halo model with gravitational dynamics.

\subsubsection{The 3-point correlation function of galaxies}
\label{gal}
It is straightforward to extend the dark matter halo approach 
to calculate the 3PCF of galaxies which can be measured
from galaxy surveys. To do this, we need
to know the mean number of galaxies per halo of given mass,
known as the halo occupation number,  $\skaco{N_{\rm gal}}(m)$, and the
second and third moments of the galaxy distribution
($\skaco{N_{\rm gal}(N_{\rm gal}-1)}$ and $\skaco{N_{\rm gal}(N_{\rm
gal}-1)(N_{\rm gal}-2)}$).  Following the method in Scoccimarro et
al. (2001), the real-space halo model expressions for the 
1-halo contributions to the galaxy 2PCF and 3PCF are given by
\begin{eqnarray}
\xi^{\rm gal}_{1h}(r)&=&2\pi \int\!\!dm~ n(m)
\frac{\skaco{N_{\rm gal}(N_{\rm gal}-1)}}{\bar{n}_{\rm gal}^2}
\int_0^{\rvir}\!\!ds
\int^\pi_0\!\!d\theta~ s^2\sin\theta~
 u_m(s)u_m(|\bm{s}+\bm{r}|),
\nonumber\\
 \zeta^{\rm gal}_{1h}(r,q,\psi)&=&\int\!\!dm~ n(m)
\frac{\skaco{N_{\rm gal}(N_{\rm gal}-1)(N_{\rm gal}-2)}}{\bar{n}_{\rm gal}^3}
\int_0^{\rvir}\!\!ds\int^{2\pi}_0\!\!d\varphi
\int^\pi_0\!\!d\theta~ s^2\sin\theta~
 u_m(s)u_m(|\bm{s}+\bm{r}|)u_m(|\bm{s}+\bm{q}|),
\end{eqnarray}
where $\bar{n}_g$ denotes the mean number density of galaxies defined as
\begin{equation}
\bar{n}_{g}=\int\!\!dm~ n(m)\skaco{N_{\rm gal}}(m).
\end{equation}
In a similar manner, we can derive halo model predictions for the
2-halo contribution to the galaxy 2PCF, and the 2- and 3-halo
contributions to the galaxy 3PCF. 
For simplicity we assume that the distribution of galaxies
within a given halo follows the dark matter profile $u_m(r)$. 
Note that in the large-scale limit, 
the galaxy bias parameter is defined as 
\begin{equation}
b_g=\frac{1}{\bar{n}_g}\int\!dm~ n(m)b(m)
\frac{\skaco{N_{\rm gal}}(m)}{m}. 
\label{eqn:gbias}
\end{equation}

The halo occupation number has been used in
the literature to explain galaxy clustering properties
(Jing, Mo \& B\"orner 1998; Seljak 2000; Peacock \& Smith 2000; 
Scoccimarro et al. 2001; Sheth et al. 2001; 
Berlind \& Weinberg 2002;  
Moustakas \& Somerville 2001; 
Scranton 2002; Cooray 2002). In
this paper, we employ the model of Scranton (2002), which approximately
reproduces the results for $\skaco{N_{\rm gal}}(m)$ from 
semi-analytic models in Kauffmann et al. (1999). 
The model gives $\skaco{N_{\rm gal}}$
for red and blue galaxies\footnote{Both samples are 
brighter than $M_V=-17.7+5\log h$ in the simulations and the red and 
blue galaxies are labelled as being redder or bluer than $B-I=1.8$.}
\begin{equation}
\skaco{N_{\rm gal}}(m)=
\left\{
\begin{array}{ll}
\left(\frac{m}{m_R}\right)^{\gamma_R}
e^{-(m_{R0}/m)^{1/2}}& \mbox{for red galaxies},\\
\left(\frac{m}{m_B}\right)^{\gamma_B}
+Ae^{-A_0\left(\log m-m_{B_s}\right)^2} &\mbox{for blue galaxies}
\end{array}
\right.
\label{eqn:occup}
\end{equation}
where the parameters $\gamma_R=1.1$, $m_R=1.8\times 10^{13}h^{-1}M_\odot$ and
$m_{R0}=4.9\times 10^{12}h^{-1}M_\odot$ for red galaxies, while
 $\gamma_B=0.93$, $m_B=2.34\times 10^{13}h^{-1}M_\odot$, $A=0.65$, 
$A_0=6.6$ and $m_{B_s}=11.73$ for blue galaxies.   In the following, 
we employ a lower mass cutoff of $m\ge 10^{11}~ h^{-1} M_\odot$ 
for the calculations, since in small halos 
effects such as supernova winds can blow away the gas from halos and
thus suppresses further star formation. 
The model above yields bias parameters $b_{g,{\rm red}}=1.51$ 
and $b_{g,{\rm blue}}=0.82$ for red and blue galaxies, 
respectively.  

\begin{figure}
  \begin{center}
    \leavevmode\epsfxsize=9.cm \epsfbox{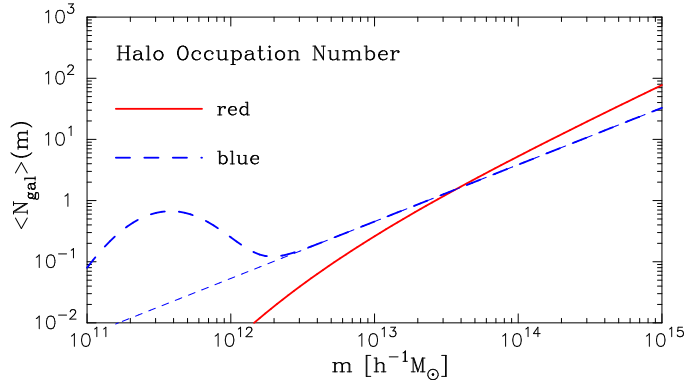}
  \end{center}
\caption{Mean number of red and blue galaxies 
as a function of parent halo mass using the forms in equation 
(\ref{eqn:occup}). The thin dashed curve shows the model for blue
 galaxies without the bump feature, 
which we will use to demonstrate the effect 
 on the halo model prediction for the 3PCF in Figure \ref{fig:qgaluv}.}
\label{fig:occup}
\end{figure}
Figure \ref{fig:occup} plots the mean number of galaxies, 
$\skaco{N_{\rm gal}}(m)$, as a 
function of parent halo mass. The figure shows that $\skaco{N_{\rm
gal}}\le 1$ at $m\simlt 10^{13}~ h^{-1}M_\odot$ for both types
of galaxies; a significant fraction of low mass halos are thus 
likely to contain at most one galaxy. For such low mass halos, 
we need to carefully model the higher order
moments of the galaxy distribution within them, as discussed below.
It is also apparent that $\skaco{N_{\rm gal}}$ 
for blue galaxies has a bump feature for halos of $m\simlt
10^{12}~ h^{-1}M_\odot$.
The bump is due to the prescriptions used in the semi-analytical model.
Although stars are formed from the cold gas in halos within a
dynamical time scale, it is assumed that the gas cooling in halos with
circular velocity $V_c>350 $ km s$^{-1}$ does not form visible stars
so that the output galaxy catalog can fit the observed Tully-Fisher
relation (see Kauffmann et al. 1999 for details).  This sharp cutoff of
the star formation prevents further formation of galaxies in halos with
$V_c>350$ km s$^{-1}$ unless the halos experience merging, which is
taken as the trigger to induce the starburst.  Note that the circular
velocity $V_c=350$km s$^{-1}$ roughly corresponds to halo of $m\sim
10^{13}M_\odot$ at the present epoch; the velocity is larger for the same
mass halo at earlier epochs.  In fact, we will show that the bump feature
drastically affects the configuration dependence and amplitude of the
reduced 3PCF of blue galaxies.

As mentioned above, we also need the second and third
moments of the galaxy distribution within parent halos, 
$\skaco{N_{\rm gal}(N_{\rm
gal}-1)}$ and $\skaco{N_{\rm gal}(N_{\rm gal}-1)(N_{\rm gal}-2)}$.  We
follow Scranton (2002), who expressed them as
$\skaco{N_{\rm gal}(N_{\rm gal}-1)
}\approx \alpha_g^2(m)\skaco{N_{\rm gal}}^2$ and
$\skaco{N_{\rm gal}(N_{\rm gal}-1)(N_{\rm gal}-2)}\approx
\alpha_g^3(m)\skaco{N_{\rm gal}}^3$, where
\begin{eqnarray}
\alpha_g(m)=
\left\{
\begin{array}{ll}
1&m>10^{13}h^{-1}M_\odot,\\
\ln(\sqrt{m/10^{11}h^{-1}M_\odot})& m<10^{13}h^{-1}M_\odot.
\end{array}
\right.
\end{eqnarray}
This model implies that galaxies follow a Poisson distribution for
massive halos $m\simgt 10^{13}h^{-1}M_\odot$ and a sub-Poisson 
distribution for smaller halos.  
Other treatments of the sub-Poisson distribution have been
proposed by Sheth et al. (2001) and Scoccimarro et al. (20001).

We expect that the models of the 2nd and 3rd moments of 
$N_{\rm gal}(m)$ will affect $Q_{\rm gal}$ at small scales which
are dominated by the 1-halo term. Large scales  $\simgt 5~ h^{-1}$Mpc,
in the quasi-linear regime, are dominated
by the 3-halo term. The behavior of $Q_{\rm gal}$ on the large scales
can be roughly expressed in terms of the bias parameter (\ref{eqn:gbias})
as $Q_{\rm gal}\sim Q_{\rm
mass}/b_{g}$. This follows from 
 $Q_{\rm gal}\sim \zeta_{\rm gal}/(3\xi_{\rm gal}^2)$ and that:
$\zeta_{\rm  gal}\sim b_{g }^3\zeta_{\rm mass}$ 
and $\xi_{\rm gal}\sim b_{g}^2\xi_{\rm mass}$.
As a result, the anti-biasing and bias of blue and red galaxies should lead to
larger or smaller amplitudes of $Q_{\rm gal}$ compared to $Q_{\rm
mass}$, respectively. 

\begin{figure}
  \begin{center}
    \leavevmode\epsfxsize=14.cm \epsfbox{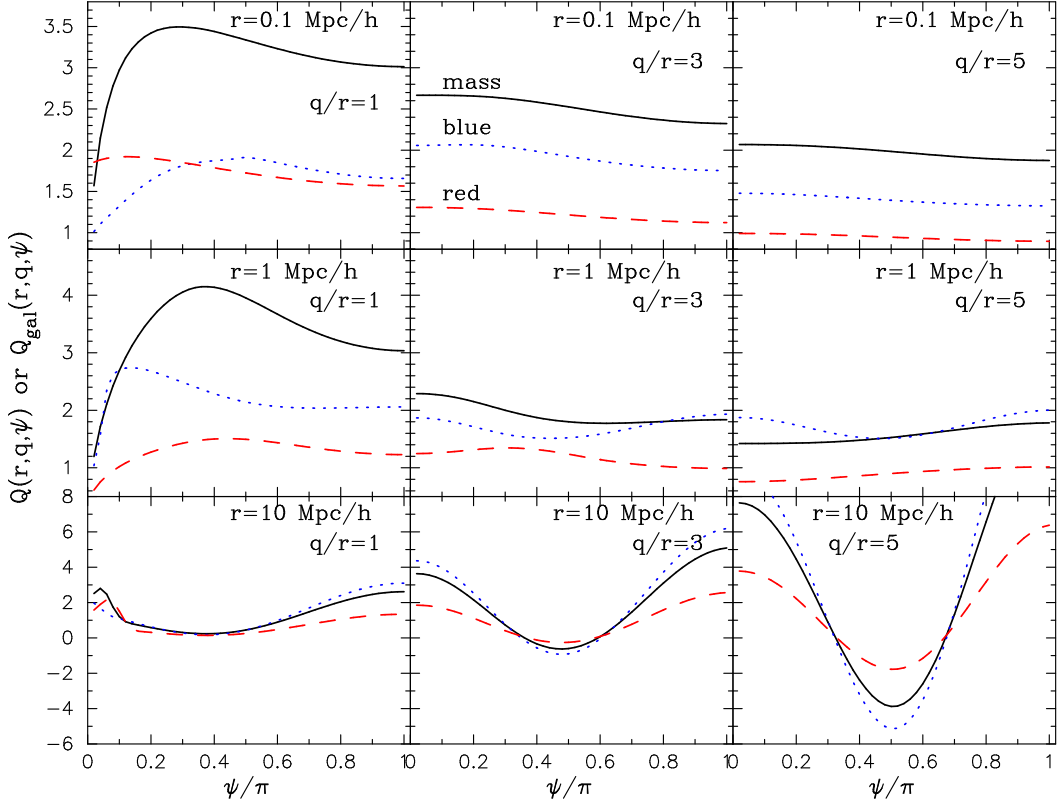}
  \end{center}
\caption{Halo model predictions for the galaxy 3PCF as a function of
triangle shapes parameterized by $r$, $q$ and $\psi$ (see Figure
\ref{fig:triang}). The solid, dashed and dotted curves show the results
for the mass and for blue and red galaxies, respectively. We have
employed the model of equation (\ref{eqn:occup}) for the the halo 
occupation number of red and blue galaxies.  } \label{fig:qgal}
\end{figure}
Figure \ref{fig:qgal} shows $Q_{\rm gal}$ for the galaxy
3PCF with triangle configurations parameterized by $r$, $q$ and $\psi$ as
in Figure \ref{fig:qpara}. For red galaxies, 
the halo model combined with the halo occupation number 
leads to a weaker configuration dependence  
and  smaller amplitude for $Q_{\rm gal}$ than the mass 3PCF, as suggested from
the measurements in Jing \& B\"orner (1998).  
On the other hand, the 3PCF of blue galaxies 
displays complex features. This is mainly due to the
bump feature in $\skaco{N_{\rm gal}}(m)$ at $10^{11}\le m\simlt
10^{13}~ h^{-1}M_\odot$, as shown below. Hence, 
a detailed
knowledge of the configuration dependence of  $Q_{\rm gal}$
could be used to quantitatively constrain the occupation number, or more
generally, the galaxy formation scenario as a function of observed
galaxy properties (morphology, luminosity and so on) within a 
halo of given mass.

\begin{figure}
  \begin{center}
    \leavevmode\epsfxsize=14.cm \epsfbox{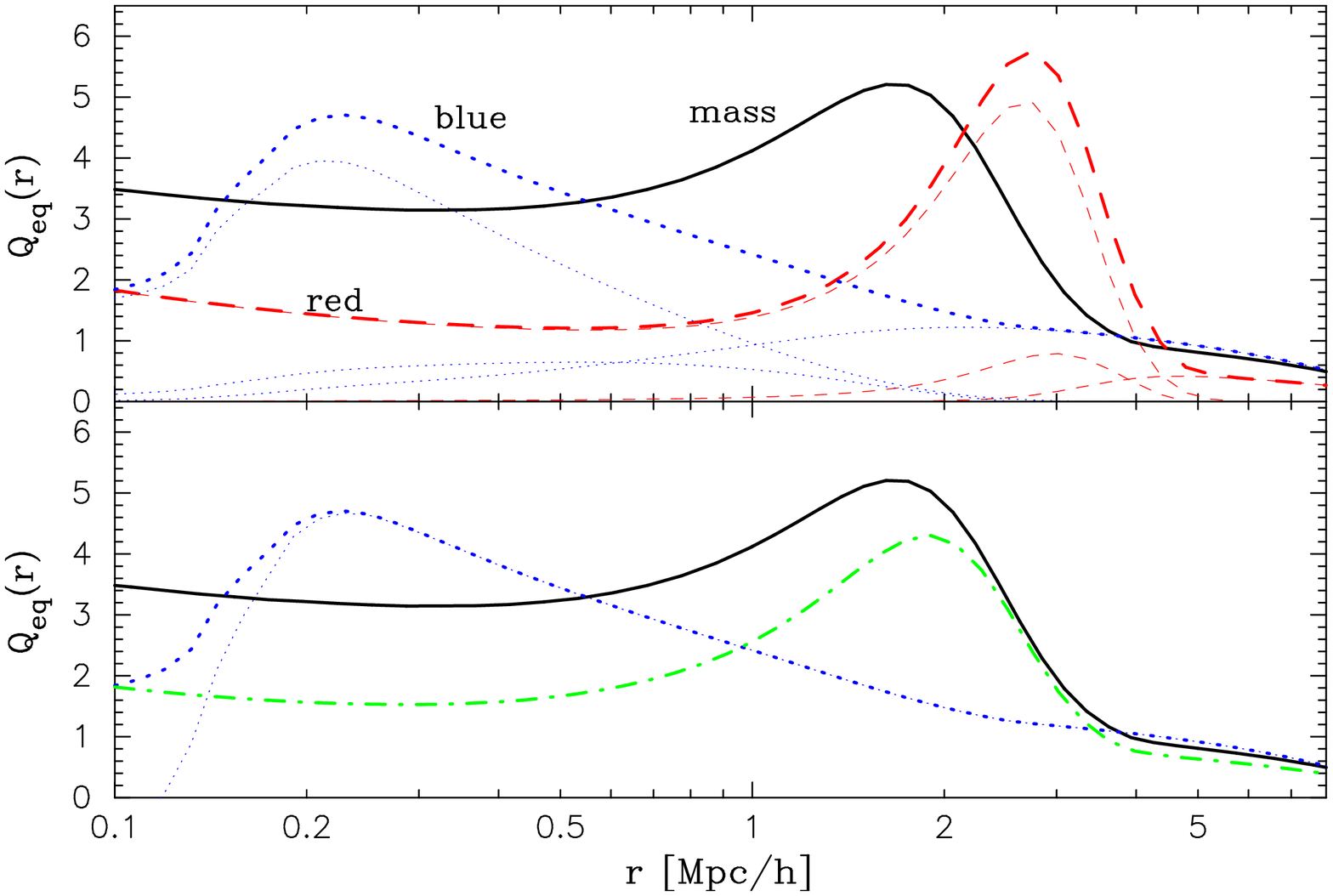}
  \end{center}
\caption{The upper panel shows 
the reduced 3PCF of red (dashed curves) and blue
(dotted curves) galaxies as a function of the side length of equilateral
triangles, as in Figure \ref{fig:qparauv}.  The thin curves show
the 1-, 2- and 3-halo contributions separately. 
For comparison, 
the solid curve shows the result for the mass 3PCF. 
Note that the bias parameters
are $b_{g,{\rm red}}=1.51$ and $b_{g,{\rm blue}}=0.82$ for the red and
 blue galaxies, respectively. The lower panel explores the dependence
of $Q_{\rm gal}$ for blue galaxies on the model ingredients. 
The dot-dashed curve 
shows the result for the $Q_{\rm gal}$ of blue galaxies, if we 
suppress the bump feature in $\skaco{N_{\rm gal}}(m)$ at 
$10^{11}\le m\simlt 10^{13}~ h^{-1}M_\odot$ for the calculation. 
The thin dotted curve shows how the model of
Scoccimarro et al. (2001) for the sub-Poisson
distribution of blue galaxies
changes the prediction at $r\simlt 0.2~ h^{-1}$Mpc.}
\label{fig:qgaluv}
\end{figure}
\begin{figure}
  \begin{center}
    \leavevmode\epsfxsize=14.cm \epsfbox{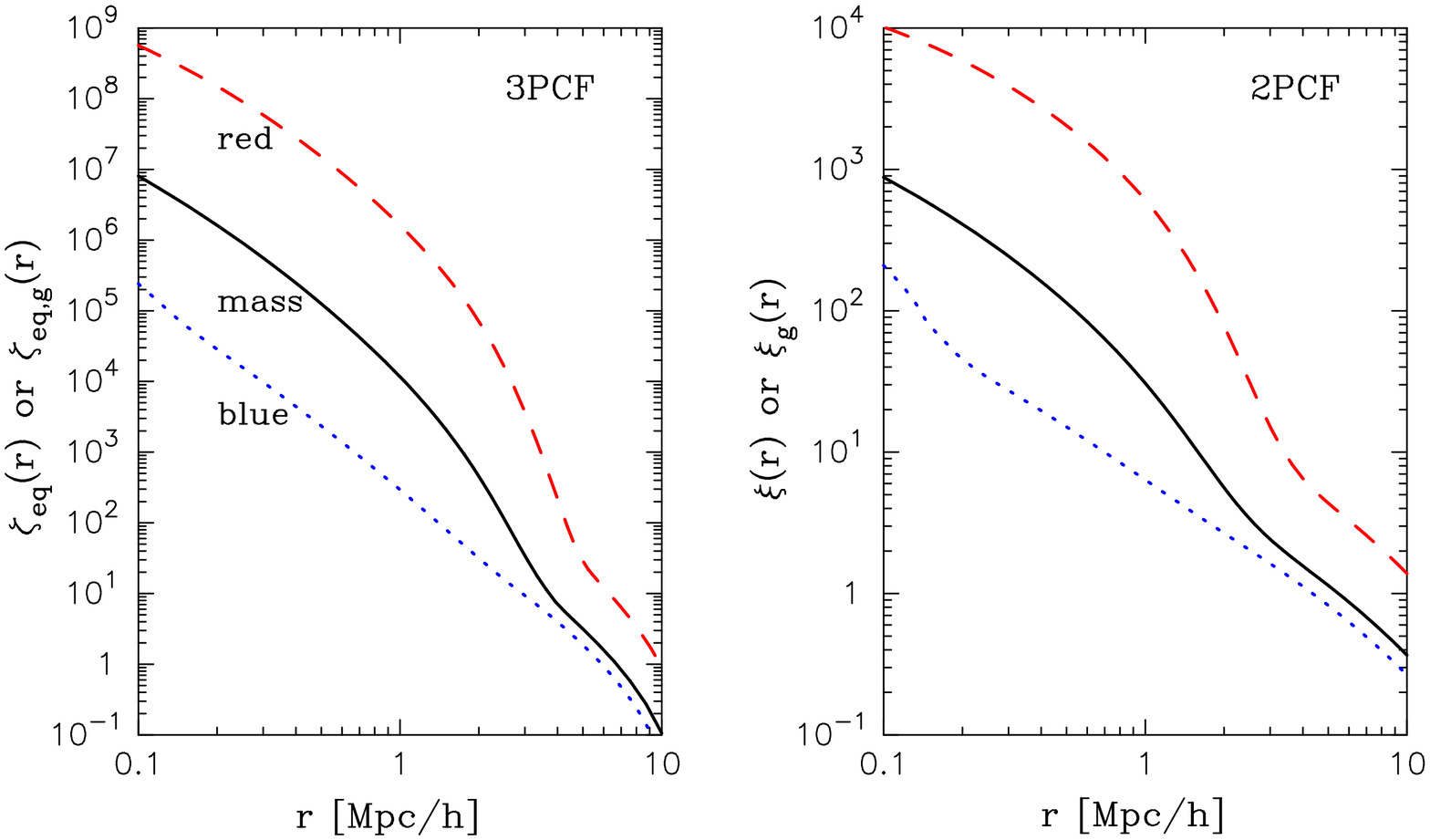}
  \end{center}
\caption{The 3PCF (left) and 2PCF (right)
of mass, red and blue galaxies.}
\label{fig:g3pt}
\end{figure}
The sensitivity of the input model of 
the halo occupation number to the galaxy 3PCF
can be more clearly understood from
Figure \ref{fig:qgaluv}, which shows the $Q_{\rm gal}$ parameters 
for equilateral
triangles. The 1-, 2- and 3-halo contributions to the galaxy 3PCF are
explicitly shown. This figure may be compared with the result 
for the galaxy bispectrum
shown in Figure 6 of Scoccimarro et al. (2001). 
It is again apparent that the amplitude of $Q_{\rm
gal}$ for red galaxies is suppressed compared to the mass 3PCF, while 
the result for blue galaxies exhibits complex features.  
Figure \ref{fig:g3pt} shows halo model predictions for 
the 2PCF and 3PCF of mass and blue and red
galaxies separately. It is worth noting that 
the 2PCF and 3PCF of blue galaxies are 
well approximated by power laws at scales $r\simgt 0.2~ h^{-1}$Mpc, 
unlike the case for the mass or for red galaxies. 
This is due to the fact that galaxy formation is inefficient in 
massive halos and thus suppresses the 1-halo term for blue galaxies 
compared to the mass or to red galaxies which are formed by mergers. 
Also, $Q_{\rm gal}$ for
blue galaxies has contributions from the 2- and 3-halo terms 
over a wider range of scales compared to that of red galaxies. This is mainly
due to the bump feature in $\skaco{N_{\rm gal}}$ at low
mass scales in Figure \ref{fig:occup}, and partly due to 
the shallower mass slope in $\skaco{N_{\rm gal}}\propto m^{\gamma}$. 

The lower panel of Figure \ref{fig:qgaluv}
explores the origin of  
the complex features in $Q_{\rm gal}$ for blue
galaxies at $0.1\simlt r\simlt 2~ h^{-1}$Mpc. 
The dot-dashed curve  shows how the halo model prediction
for  $Q_{\rm gal}$ for blue galaxies changes if we suppress the 
bump feature in $\skaco{N_{\rm gal}}$ 
at low mass scales  by setting  $A=0$ in equation (\ref{eqn:occup}).
The model $\skaco{N_{\rm gal}}$ used is shown by the thin dashed curve
in Figure \ref{fig:occup}.
One can see that the complex features in $Q_{\rm gal}$ disappear,
except for the problematic bump feature at $r\sim 2$Mpc (see the discussion
around Figure \ref{fig:excl}), and its amplitude becomes smaller than
the mass 3PCF because of the change of the bias parameter to $b_{\rm
g}=1.24$ from $b_{\rm g}=0.82$, which results in behavior similar to the red
galaxies. Thus, the input model of $\skaco{N_{\rm
gal}}$ has a drastic effect on $Q_{\rm gal}$.  We also investigate how
the halo model predictions are affected by altering the sub-Poisson
distribution for the galaxy third-order moment, which is relevant for
the range $\skaco{N_{\rm gal}}\le 1$ or equivalently $m\simlt 10^{13}~
h^{-1}M_\odot$.  Scoccimarro et al. (2001) proposed the binomial
distribution, where the third-order moments is given by $\skaco{N_{\rm
gal}(N_{\rm gal}-1)(N_{\rm
gal}-2)}=\alpha_g^2(2\alpha_g^2-1)\skaco{N_{\rm gal}}^3$. Note that this
model leads to negative values for the third moment for $\alpha_g\ll 1$.
The thin dotted curve in Figure \ref{fig:qgaluv} 
shows the result of using this prescription --- it
changes $Q_{\rm gal}$ only for scales $r\simlt 0.2~ h^{-1}$Mpc.
 
\subsection{The 3-point correlation function of weak lensing fields}

Next we consider the halo model predictions for the angular 3PCF of the
weak lensing convergence field. It should be noted that the following
method can be easily extended to the predict the angular 3PCF 
of the galaxy distribution, as discussed in \S \ref{ang3pt}.

The 1-halo term for the 3PCF of $\kappa$ is given by
the following 4-dimensional integral: 
\begin{eqnarray}
Z_{1h}(r,q,\psi)=
\int^{\chi_s}_{0}\!\!d\chi~ d_A^2(\chi)
\int\!\!dm~ n(m; \chi)
\int^{\theta_{\rm vir}}_0\!\!d\theta\int^{2\pi}_0\!\!d\varphi~ 
\theta\ \kappa_m(\theta)\kappa_m(|\bm{\theta}+\bm{r}|)
\kappa_m(|\bm{\theta}+\bm{q}|),
\end{eqnarray}
where $|\bm{\theta}+\bm{r}|=(\theta^2+r^2 +2\theta r\cos\psi)^{1/2}$ and
$|\bm{\theta}+\bm{q}|=(\theta^2+q^2 +2\theta
q\cos(\varphi-\psi))^{1/2}$. For the NFW profile, $\kappa_m$ is analytically
given by equations (\ref{eqn:formsig}) and (\ref{eqn:kappam}).
Note that $r$ and $q$ denote angular
scales, although we use the same notations as for the 3D case for
simplicity.  In analogy with equation (\ref{eqn:qpara}), the
$Q$ parameter for the angular 3PCF is defined as
\begin{equation}
Q_\kappa(r,q,\psi)=
\frac{Z(r,q,\psi)}
{w_\kappa(r)w_\kappa(q)+w_\kappa(r)
w_\kappa(|\bm{r}-\bm{q}|)+w_\kappa(q)w_{\kappa}(|\bm{r}-\bm{q}|)}, 
\end{equation}
where $Z=Z_{1h}+Z_{2h}+Z_{3h}$. Note that the 2- and 3-halo terms,
$Z_{2h}$ and $Z_{3h}$, are computed from the approximations of equations
(\ref{eqn:app2d2h}) and (\ref{eqn:app2d3h}), respectively.
 
\begin{figure}
  \begin{center}
    \leavevmode\epsfxsize=17.cm \epsfbox{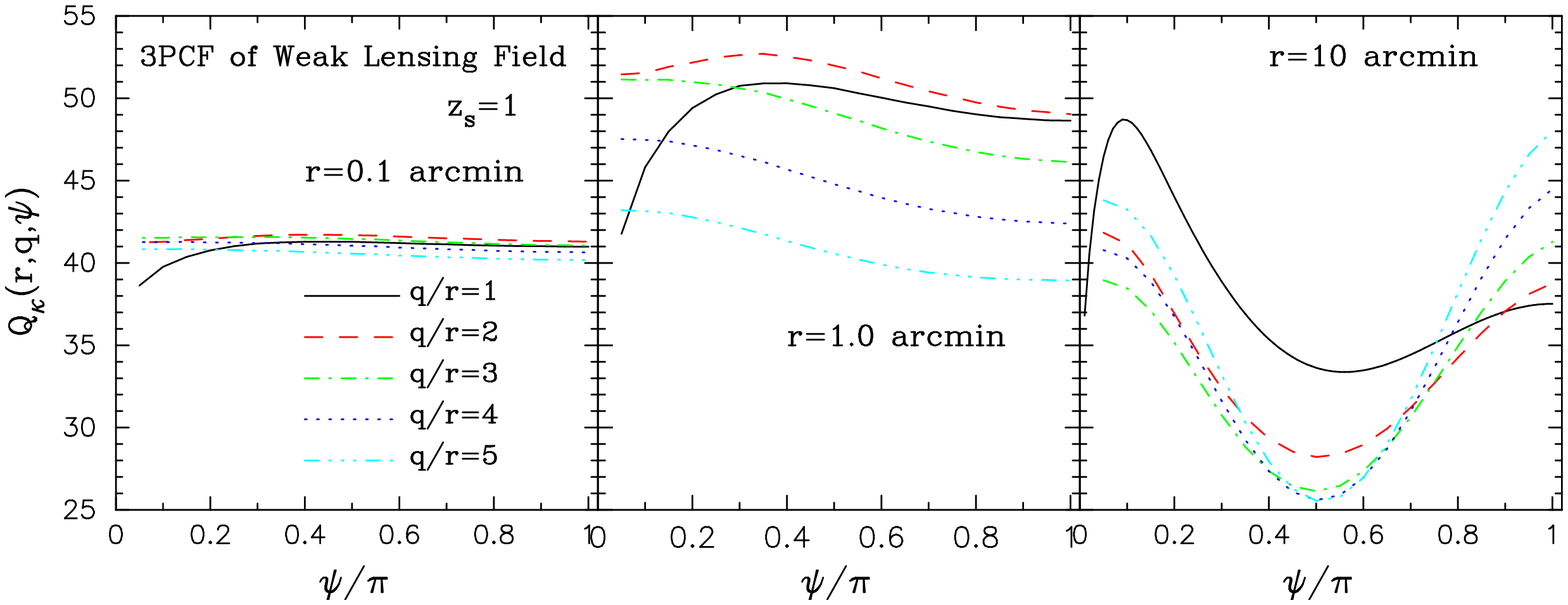}
  \end{center}
\caption{Halo model predictions for the reduced $Q$
parameter for the weak lensing convergence, with triangle 
configurations parameterized as in Figure \ref{fig:qpara}. 
The configuration dependence of $Q$ changes as one shifts 
from the strongly nonlinear regime shown in the left panel
to the quasi-linear regime shown in the right panel.  
} 
\label{fig:AngQ}
\end{figure}
\begin{figure}
  \begin{center}
    \leavevmode\epsfxsize=14.cm \epsfbox{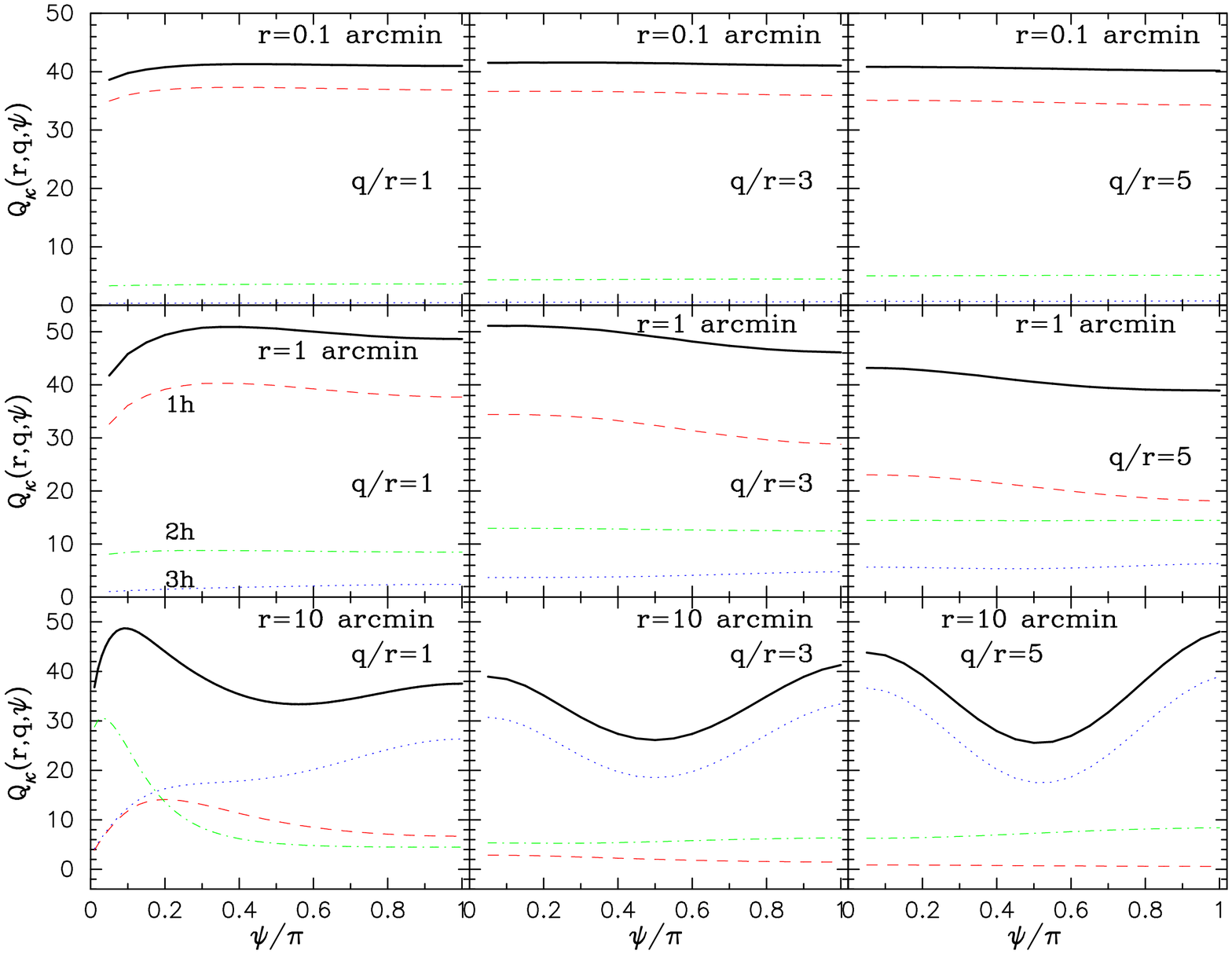}
  \end{center}
\caption{The 1-, 2- and 3-halo contributions to the 
$Q$ parameter are shown for the convergence, with triangle 
shapes parameterized as in Figure \ref{fig:AngQ}. 
}
\label{fig:AngQ1h}
\end{figure}

Figure \ref{fig:AngQ} shows halo model predictions for $Q_{\kappa}$
with triangle configurations parameterized as in Figure 
\ref{fig:qpara}. The halos are taken to have NFW profiles. 
The hierarchical ansatz, $Q_{\kappa}={\rm const.}$, does not hold rigorously
over the scales we have considered, as in the 3D case. 
However, the configuration dependence of 
$Q_{\kappa}$ for $r=0.1'$ is much weaker than for $r=1'$ and $10'$.  
The strong configuration dependence in the two panels on the 
right can be understood as follows. 
First, the results in Figure \ref{fig:qpara}
show that the $Q$ parameter for the 3D mass distribution has a 
stronger configuration dependence for $r \simgt 1~ h^{-1}$Mpc
than on strongly non-linear scales.  The weak lensing convergence is
a weighted projection of the mass distribution. As a result 
of the lensing projection, a range of 3-dimensional length scales
contributes to
$Q_{\kappa}$ at a given $\theta$, which is therefore more sensitive 
to the 2- and 3-halo terms than the 3D case.   
This is explicitly shown in Figure \ref{fig:AngQ1h}, which
plots the 1- and 2- and 3-halo contributions to 
$Q_{\kappa}$ for the triangle shapes in Figure
\ref{fig:AngQ}. For the smallest scale, 
$r=0.1'$, the 1-halo term indeed yields the dominant
contribution to $Q_{\kappa}$, but the 2-halo term still
has a non-negligible contribution ($\sim 10\%$). 
For $r=1'$, the 2-halo term becomes important ($\simgt 20\%$ 
contribution).  The 3-halo term becomes dominant at
for $r \simgt 10'$, at which $Q_{\kappa}$ displays a characteristic,
oscillatory feature as predicted by perturbation
theory. 

\begin{figure}
  \begin{center}
    \leavevmode\epsfxsize=14.cm \epsfbox{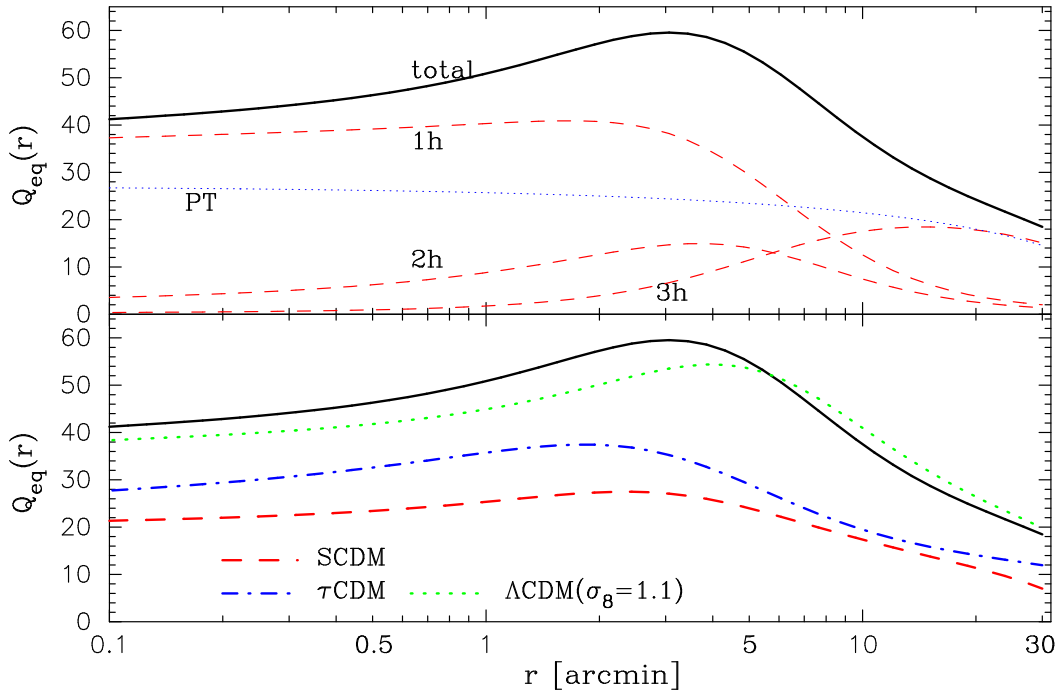}
  \end{center}
\caption{The $Q$ parameter for the convergence for equilateral
triangles as a function of the side length $r$ (in arcminutes), as in
 Figure \ref{fig:qparauv} and \ref{fig:qcosmo}. 
}
\label{fig:AngQuv}
\end{figure}
In Figure \ref{fig:AngQuv} we show $Q_\kappa$  for
equilateral triangles as a function of side length.  One can readily
see that projection effects lead to greater contributions from the
2- and 3-halo terms, in contrast to
the 3D case shown in Figure \ref{fig:qparauv}. One caveat that 
should be noted, as discussed for Figure \ref{fig:qparauv} and 
\ref{fig:excl}, is that
the halo model might overestimate $Q$
 at the transition scale between the quasi-linear and non-linear
regimes. This would be reflected in $Q_{\kappa}$ 
over the angular scales $0.1'\simlt r\simlt 20'$,
where the 2-halo term is relevant.  This uncertainty
needs further investigation by comparison  with
ray-tracing simulation results.  The figure also shows that, on very small
scales $\simlt 1'$, the 1-halo term of $Q$ decreases slightly
with decreasing angular scale. The small scale slope of the
1-halo term is determined by the combined effects of 
the NFW profile, the mass function and the halo
concentration in the halo model calculation. For
lensing, the small scale slope of the 2PCF and 3PCF is not transparent
due to projection effects. 

In the lower panel of Figure \ref{fig:AngQuv}, we plot 
$Q_{\kappa}$ for different
cosmological models as shown in Figure \ref{fig:qcosmo}. 
$Q_{\kappa}$ is sensitive to cosmological parameters, in particular 
to $\Omega_{\rm m0}$ parameter, in analogy with the sensitivity 
of the skewness parameters of
weak lensing to $\Omega_{\rm m0}$ 
(see e.g. Bernardeau et al. 2003; TJ02).  Hence, it
is expected that measuring $Q_{\kappa}$ can be used to break
the degeneracies in the determinations $\Omega_{\rm m0}$ and $\sigma_8$
so far from measurements of the shear 2PCF (e.g., Van Waerbeke et al. 2001).

\subsection{Performance of our approximation for calculating the 1-halo term of the 3PCF}
\label{appperform}

\begin{figure}
  \begin{center}
    \leavevmode\epsfxsize=13.cm \epsfbox{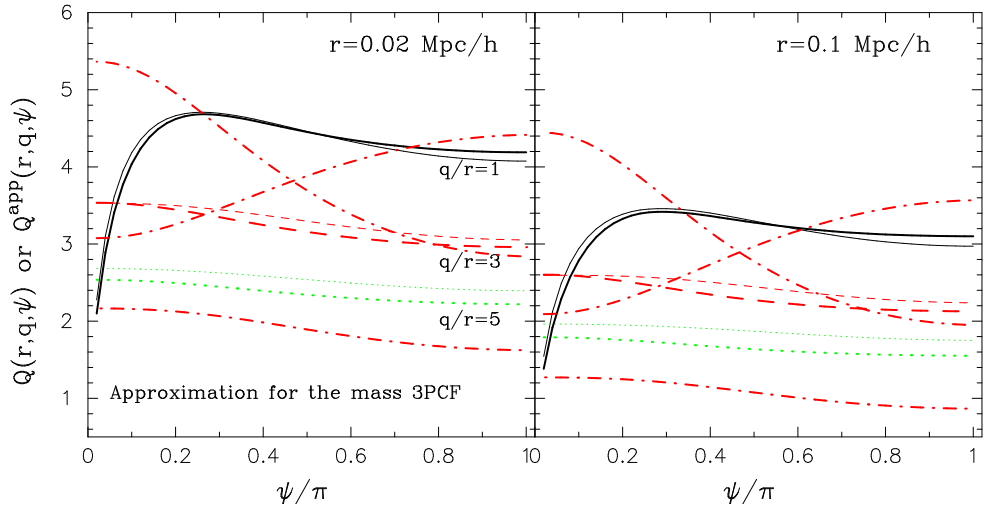}
  \end{center}
\caption{The performance of our approximation (\ref{eqn:app1h}) for
calculating the 1-halo contribution to the mass 3PCF. The results are
shown for the triangle configurations as in Figure \ref{fig:qpara}. 
We consider the non-linear scales $r=0.02$ (left panel) and
$0.1~ h^{-1}$Mpc (right). The thin curves are the exact halo model
predictions, while the thick curves show the results of the approximation
of equation (\ref{eqn:app1h}).  The three dot-dashed curves show each
term in the approximation (\ref{eqn:app1h}) for 
$q/r=3$. This figure verifies that our approximation 
accurately describes the 3PCF, 
even though each term in equation (\ref{eqn:app1h}) is
far from the exact value.  } \label{fig:appq}
\end{figure}

In Appendix \ref{app} we present a useful approximation for calculating the
one halo contribution to the mass 3PCF, which uses
the Fourier-space halo model combined with the
approximation developed in TJ02 for calculating the skewness and
kurtosis parameters.  
Figure \ref{fig:appq} demonstrates the performance of our approximation
by comparing the results with the exact values computed from the
real-space halo model. We have used the approximation for the 3PCF
calculation in the numerator of $Q$, and the 2PCF calculation in the
denominator includes full contributions from the 1- and 2-halo terms.
We consider some of the triangle shapes shown in Figure
\ref{fig:qpara}. The left and right panels show the results for $r=0.02$
and $0.1~ h^{-1}{\rm Mpc}$. In each panel, the solid, dashed and dotted
curves are for $q/r=1$, $3$ and $5$, respectively.  
Note that all the triangles considered are
in the strongly non-linear regime since we wish to check the validity of our
approximation for the 1-halo term.  The thin curves are the exact
predictions from the real-space halo model computed from equation
(\ref{eqn:r3ptcalc}), while the thick curves denote the results of the
approximation (\ref{eqn:app1h}). The comparison demonstrates that the
approximation works remarkably well with an accuracy of $\simlt 5\%$ for
 $q/r=1$ and $3$, and $\simlt 10\%$ even for elongated triangles with 
$q/r=5$.  The three dot-dashed
curves show the results of each term in the approximation
(\ref{eqn:app1h}) for $q/r=3$. Interestingly, although each
term does not work well, the approximation obtained by summing 
them is close to the exact value.  We have confirmed that the
approximation of equation (\ref{eqn:2dapp1h}) for the 1-halo term 
of the weak lensing 3PCF also works to about 
the same accuracy as the 3D case shown here.

\section{Discussion and Conclusion}
\label{discuss}
In this paper, we have used the halo clustering model
to compute the 3-point correlation function (3PCF) 
of cosmological fields. We have shown results for
the three-dimensional mass and galaxy distributions and the
two-dimensional weak lensing convergence field.  The halo model enables us to
separately consider the contributions to the 3PCF arising from 
triplets in a single halo or two or three different halos.  Thus
we can understand how gravitational clustering transitions from the 
quasi-linear
regime to the strongly non-linear regime as one goes to smaller scales. 
We found that the single halo contribution,
which is dominant on small scales, can be computed 
using the real-space formulation of the halo model far more easily
than the Fourier-space approach used in the literature.  We 
also developed approximations for
computing the 2- and 3-halo contributions to the 3PCF.  Since
measuring the real space 3PCF on small scales 
is  likely to be easier than the bispectrum, 
our model predictions will allow 
for the extraction of cosmological information from forthcoming
galaxy surveys and weak lensing surveys. 

We obtained the following results by applying our halo model to 
the concordance CDM model ($\Lambda$CDM). The 
3PCF on small scales $r\simlt 1~ h^{-1}{\rm Mpc}$ is dominated
by the 1-halo term. Hence it probes properties of massive halos
and can be used to constrain halo profiles as discussed below. 
The quasi-linear
3PCF predicted by perturbation theory can be reproduced by the
3-halo contribution for $r\simgt 3~ h^{-1}$Mpc.  
For plausible halo model parameters, the hierarchical ansatz for the
reduced 3PCF parameter, $Q= {\rm constant}$, does not hold over the
range of scales we have considered. However the $Q$ parameter does have
a weaker dependence on  triangle configurations in the non-linear
regime than in the quasi-linear regime (see Figures \ref{fig:qpara} and
\ref{fig:qjing}).  These results 
qualitatively verify the results in Ma \& Fry (2000b,c). 
Ma \& Fry also pointed out that the halo model for plausible
model parameters violates the stable clustering
hypothesis which has been widely used to develop analytical prediction
of the non-linear gravitational clustering, 
as done in the popular PD fitting formula.  In fact, very recently Smith et
al. (2002) showed a weak violation of the stable clustering hypothesis
using high-resolution $N$-body simulations.  However,
our halo model predictions for the 3PCF do not match the
asymptotic shape proposed in Ma \& Fry (2000b) because of the
assumptions employed in the asymptotic formula.  Therefore, it
it merits further study to carefully investigate how small-scale gravitational
clustering can be described by the halo model ingredients. Such a study
can be carried out with the halo model methods developed in this paper. 

We also found that the non-linear 3PCF is most sensitive to the inner slope of
the halo profile and to the halo concentration parameter for given 
cosmological parameters (see
Figures \ref{fig:qprofr001}, \ref{fig:qprofr01} and \ref{fig:qalpha}).  
Combinations of the inner slope and the concentration can be
adjusted so that the halo model matches the 2PCF result 
(see Figure \ref{fig:2ptdep}).  However only one combination can match
both the 2- and 3PCFs. Hence in combination with the 2PCF, the non-linear 
3PCF could be used to 
constrain the properties of dark matter halos. We suggest that the
use of higher order correlations is a useful way of measuring the
properties of massive halos. For example, while lensing surveys of clusters
can be used to measure cluster halo profiles directly, this involves 
identifying cluster centers and assigning masses to clusters to measure
averaged profiles. In contrast, by measuring the 2- and 3PCFs, parameters
of the halo mass function and profile can be fitted for. While this approach
is more challenging computationally, and requires some theoretical 
assumptions, it treats the data more objectively. 

Our halo model predictions match the simulation results of Jing \&
B\"orner (1998) in the non-linear regime as well as the quasi-linear
regime, as shown in Figure \ref{fig:qjing}.
However, the halo model seems to be less accurate on 
the intermediate transition scale $\sim 1~h^{-1}{\rm Mpc}$.  Figure
\ref{fig:qparauv} shows that the predicted $Q$ parameter for equilateral
triangles has a bump  at this scale, which corresponds to
the bump found for the reduced bispectrum at $k\sim 1~ h\ {\rm
Mpc}^{-1}$ by Scoccimarro et al. (2001). 
This is unlikely to reflect real properties of dark matter
clustering. Rather, we argued that the bump feature is sensitive
to the sharp cutoff of halos at the virial radius and the exclusion 
effect for the 2- and 3-halo terms. We explored alternative 
prescriptions as shown in Figure \ref{fig:excl}. 
To resolve this, detailed calculations and comparisons
of the halo model predictions with simulation results will be needed. 

We extended the halo model to predict
the 3PCF of the galaxy distribution.  Once we model
how galaxies populate a halo of given mass, the halo
occupation number, we can straightforwardly predict the galaxy 3PCF
based on the halo model developed here. For the halo occupation number 
for red galaxies we have used, the galaxy 3PCF has a smaller 
amplitude and weaker dependence on
triangle configuration compared to the mass 3PCF (Figures
\ref{fig:qgal} and \ref{fig:qgaluv}).  This trend is indeed consistent
with the actual measurements in Jing \& B\"orner (1998).  On the
other hand, the 3PCF of blue galaxies displays complex features
reflecting the input model of the halo occupation number. 
Thus, the galaxy 3PCF can be used to constrain the galaxy formation 
scenario as a function of host halo properties and  galaxy type. Further
work is needed to model the expected properties of galaxies by
type for specific surveys and compute the resulting 3PCF. 

As an example of the angular 3PCF, we have computed 
the 3PCF of the weak lensing convergence field. In particular,
we employed the real-space halo model to compute the single halo
contribution, as for the 3D case, which enabled us to
compute the 1-halo term by a 4-dimensional integration. 
We verified that the real-space halo model is
equivalent to the Fourier-space model well studied in the literature
(see Figure \ref{fig:ang2pt}).  We also developed approximations for
calculating the 2- and 3-halo terms. Because of projection effects,
the 2- and 3-halo terms contribute to the 3PCF over a
wider range of scales for the angular 3PCF compared to the 3D 3PCF. 
The resulting 3PCF does not  obey
the hierarchical ansatz over the angular scales we have considered.  The
 lensing 3PCF is sensitive to 
cosmological parameters, in particular $\Omega_{\rm m0}$. 
Comparing measurements with model predictions  of the 3PCF can
be a useful tool to break degeneracies between the power spectrum and
$\Omega_{\rm m0}$.   
We intend to compute 3-point functions of the
shear field following Schneider \& Lombardi (2002) and Zaldarriaga
\& Scoccimarro (2002) as these are easier to measure from the data. 
In Takada \& Jain (2003), we presented a brief investigation of 
the 3PCFs of shear fields, where we found good agreement between 
the halo model prediction and the results measured from the 
ray-tracing simulations. 
It is also expected that the $n$-point correlations of weak lensing on 
sub-arcminute scales contain a
wealth of information on properties of massive halos, beyond
their dependence on cosmological parameters.  
The real-space halo model we have developed will be a useful 
analytical tool for such calculations. 

The halo model presented in this paper allows for several interesting
applications. First, the model can be extended to investigate the
effect of triaxial halo shapes on the 3PCF, since it is the
lowest-order statistical probe of non-sphericity.  So far halo
model applications assume a single, spherically symmetric profile.
Recently, based
on high-resolution simulations, Jing \& Suto (2002) showed that
halos are more accurately described by  triaxial halo
profiles than the spherically symmetric NFW profile. 
They claimed that the axis ratios typically
have vales of  $\sim 0.6$ for the smallest and largest axis.  
The non-sphericity of the
halo profile could lead to a characteristic configuration dependence of
the 3PCF.  This effect should also affect the interpretation of 
cosmic shear measurements (Jing 2002). Likewise, 
the formulation developed in Sheth \& Jain (2002) can be used to 
discuss the effects of substructure within halos on the 2PCF and  3PCF. 

The real-space halo model can be directly used to compute the covariance
of the 2PCF.  
As discussed in Cooray \& Hu (2001b), the mass distribution on
small scales displays pronounced non-Gaussian features induced by 
non-linear gravitational clustering. Hence one needs to take into
account the non-Gaussian errors arising from the 4-point correlation
function (4PCF).  On the  scales of interest the single halo contribution
should dominate the error. The real-space halo model
allows us to analytically compute the error contribution arising from
the 4PCF with no additional computational effort than for the 3PCF 
(see \S \ref{rhalo}).

In Appendix \ref{app}, we constructed an approximation for calculating
the 1-halo term of the 3PCF based on the Fourier-space halo model. 
We showed that the the approximation accurately
describes the amplitude and configuration dependence of the
3PCF (see Figure \ref{fig:appq}).  
We will employ this approximation to perform an analytical study of the
pairwise peculiar velocity dispersion (PVD) within the BBGKY hierarchy
formalism (Peebles 1980). The PVD can be measured through
the redshift distortions inferred from galaxy surveys (e.g.,
Zehavi et al. 2002).  The BBGKY picture tells us that the PVD arises
mainly from the 3PCF on small scales.  However,
there has been no comprehensive analytical model to describe the
non-linear PVD.  This is because of the complex form of the BBGKY
hierarchy equations. Peebles (1980) (see Mo, Jing \& B\"orner 1997;
Jing et al. 1998 for a 
detailed study) assumed the hierarchical form for the non-linear 3PCF,
although it turns out to be violated for CDM models. The isothermal
assumption for the velocity distribution within a given halo was
employed to analytically obtain the PVD based on the halo model (Sheth
et al. 2001).  It is important to 
clarify whether the BBGKY hierarchy leads to a self-consistent
PVD for the CDM model. Moreover, we can easily
combine the halo model prediction with models of the halo occupation
number of galaxies to predict the PVD for different galaxy types, 
following the approach  in \S \ref{gal}.

\bigskip

We would like to thank R. Sheth, R. Scoccimarro, I. Szapudi, 
A. Taruya and L. Hui for several valuable discussions. We thank 
Y. P. Jing for kindly providing us with his simulation data results.
Helpful comments from the referee, Chung-Pei Ma, led to improvements in 
the paper. This work is supported by NASA grants NAG5-10923, NAG5-10924 
and a Keck foundation grant.

\appendix
\section{Approximation for the 2- and 3-halo terms of the 3PCF}
\label{app3pt}

In this appendix, we present approximations used for the predictions of
the 2- and 3-halo contributions to the 3PCF of the density field. The
approximations are based on the Fourier-space halo model,
combined with an approximation for the configuration dependence
of the bispectrum from Scoccimarro et al. (2001; see also TJ02).

\subsection{3D 3-point correlation function}
Following the Fourier-space halo model, as discussed in \S \ref{fhalo}, the
2-halo term for the mass 3PCF  can be expressed
from equations (\ref{eqn:bisp}) and (\ref{eqn:f3pt}) as
\begin{eqnarray}
\zeta_{2h}(\bm{r}_1,\bm{r}_2,\bm{r}_3)
&=&\left[\int\!\!dm n(m)\left(\frac{m}{\bar{\rho}_0}\right)^2 b(m)
\int\!\!\frac{d^3\bm{k}_1}{(2\pi)^3} \tilde{u}_{m}(k_1)
\exp[i\bm{k}_1\cdot \bm{r}_{12}]\right]\tilde{u}_m(k_{12})
\nonumber\\
&&\times \left[\int\!\!dm'~ n(m')\left(\frac{m'}{\bar{\rho}_0}\right)
b(m')\int\!\!\frac{d^3\bm{k}_2}{(2\pi)^3}
\tilde{u}_{m'}(k_2)P^L(k_2)\exp[i\bm{k}_2\cdot\bm{r}_{32}]\right]
+\mbox{\rm cyc.} , 
\end{eqnarray}
where the bias parameter $b(m)$ is given by equation
(\ref{eqn:bias}). 
As discussed in Scoccimarro et al. (2001), the
physical meaning of the 2-halo term tells us that 
the main contribution to the first term on the r.h.s arises from
modes with $k_1\gg k_2$, leading to the approximation $k_{12}\approx
k_1$. Replacing $\tilde{u}_{m}(k_{12})$ with $\tilde{u}_{m}(k_1)$
allows us to perform analytically the angular integration 
in $\int d^3\bm{k}_i$
in the first term above.  Using similar procedures for the second and
third terms on the r.h.s. yields the following approximation:
\begin{eqnarray}
\zeta^{\rm app}_{2h}(\bm{r}_1,\bm{r}_2,\bm{r}_3)
&=&\left[\int\!\!dm~ n(m)\left(\frac{m}{\bar{\rho}_0}\right)^2 b(m)
\int\!\!\frac{k_1^2dk_1}{2\pi^2} (\tilde{u}_{m}(k_1))^2j_0(k_1r_{12})
\right]\nonumber\\
&&\times \left[\int\!\!dm'~ n(m')\left(\frac{m'}{\bar{\rho}_0}\right)
b(m')\int\!\!\frac{k^2_2dk_2}{2\pi^2}
\tilde{u}_{m'}(k_2)P^L(k_2)j_0(k_2 r_{23})\right]+\mbox{\rm cyc.}
\label{eqn:app2h}
\end{eqnarray}
where we have used the formula $\int\!\!d\Omega_{\bm{k}}
\exp[i\bm{k}\cdot\bm{r}]=j_0(kr)$. The large square bracket is meant to
show explicitly that the integrations inside each bracket can be done
separately from the other.  Thus one needs to perform only 
2-dimensional integrations to get the 2-halo term. It was shown
that a similar approximation for the skewness calculation turns out 
to be accurate (Scoccimarro et al. 2001; TJ02). Hence, we expect that
this also holds for the 3PCF calculation.
 
The 3-halo contribution to the 3PCF can be expressed as
\begin{eqnarray}
\zeta_{3h}(\bm{r}_1,\bm{r}_2,\bm{r}_3)&=&
\int\!\!dm~ n(m)\left(\frac{m}{\bar{\rho}_0}\right)b(m)
\int\!\!dm'~ n(m')\left(\frac{m}{\bar{\rho}_0}\right)b(m)
\int\!\!dm^{\prime \prime}~ n(m^{\prime\prime})
\left(\frac{m^{\prime\prime}}{\bar{\rho}_0}\right)b(m^{\prime\prime})
\nonumber\\
&&
\times \left[\int\!\!\frac{d^3\bm{k}_1}{(2\pi)^3}
\frac{d^3\bm{k}_2}{(2\pi)^3}u_{m}(k_1)u_{m'}(k_2)u_{m^{\prime\prime}}(k_{12})
e^{i\bm{k}_1\cdot\bm{r}_{13}+i\bm{k}_2\cdot\bm{r}_{23}}
P^L(k_1)P^L(k_2)\right.\nonumber\\ 
&&\times \left. 
\left\{\frac{10}{7}+\left(\frac{1}{k_1^2}+\frac{1}{k_2^2}
\right)\bm{k}_1\cdot\bm{k}_2+\frac{4}{7}\frac{(\bm{k}_1\cdot\bm{k}_2)^2}
{k_1^2k_2^2}\right\}+\mbox{\rm cyc.}\right].
\end{eqnarray}
We again employ the approximation  
$\tilde{u}_{m^{\prime\prime}}(k_{12})\approx
\tilde{u}_{m^{\prime\prime}}(k_1)$ to
perform analytically the integration for the first term on the r.h.s., 
although there is no guarantee that the approximation is as accurate as for the
2-halo term. 
Following the procedure developed to derive the perturbation theory 3PCF
(Fry 1984b; also see Jing \& B\"orner
1997; Gazta\~naga \& Bernardeau 1998; Barriga \& Gazta\~naga 2002)
we can derive the approximation for the 3-halo term:
\begin{eqnarray}
\zeta^{3h}_{\kappa}(\bm{r}_1,\bm{r}_2,\bm{r}_3)
&=&
\frac{10}{7}\Xi^{(2)}(r_{13})\Xi^{(1)}(r_{23})
-\frac{4}{7}
\left(-3\frac{\Phi^{(2)\prime}(r_{13})}{r_{13}}
\frac{\Phi^{(1)\prime}(r_{23})}{r_{23}}-
\frac{\Xi^{(2)}(r_{13})\Phi^{(1)\prime}(r_{23})}{r_{23}}
-\frac{\Xi^{(1)}(r_{23})\Phi^{(2)\prime}(r_{13})}{r_{13}}\right) \nonumber\\
&&
-\mu_{1323}\left(
\Xi^{(2)\prime}(r_{13})\Phi^{(1)\prime}(r_{23})
+\Xi^{(1)\prime}(r_{23})\Phi^{(2)\prime}(r_{13})\right)\nonumber\\
&&+\frac{4}{7}\mu_{1323}^2
\left(\Xi^{(2)}(r_{13})+3\frac{\Phi^{(2)\prime}(r_{13})}{r_{13}}\right)
\left(\Xi^{(1)}(r_{23})+3\frac{\Phi^{(1)\prime}(r_{23})}{r_{23}}\right)+{\rm cyc. },
\label{eqn:app3h}
\end{eqnarray}
where $\mu_{1323}\equiv (\bm{r}_{13}\cdot \bm{r}_{23})/(r_{13}r_{23})$ and 
\begin{eqnarray}
\Xi^{(\mu)}(r)&\equiv& \int\!\!\frac{k^2dk}{2\pi^2 } \left[\int\!\!dm~ 
n(m)\left(\frac{m}{\bar{\rho}_0}\right) b(m)u_m(k)\right]^\mu 
P^L(k)\frac{\sin(kr)}{kr}, \nonumber \\
\Phi^{(\mu)}(r)&\equiv &\int\!\!\frac{k^2dk}{2\pi^2 } \left[\int\!\!dm~ 
n(m)\left(\frac{m}{\bar{\rho}_0}\right) b(m)u_m(k)\right]^\mu 
\frac{P^L(k)}{k^2}\frac{\sin(kr)}{kr}. \nonumber \\
\end{eqnarray}
We find that this approximation reproduces the perturbation theory result 
at scales $\simgt 5~ h^{-1}$Mpc for the \LCDM model (see Figure
\ref{fig:qparauv}). Note that, for the actual
predictions of the 3PCF shown in this paper,  we have used
the permutation $(1)\leftrightarrow(2)$ for terms such as
$\Xi^{(2)}\Xi^{(1)}$ in the above equation, so that it satisfies 
statistical symmetry for permutations between $\bm{r}_1$, $\bm{r}_2$ and
$\bm{r}_{3}$. 

By replacing $(m/\bar{\rho}_0)$ in equations (\ref{eqn:app2h}) and 
(\ref{eqn:app3h}) with the halo occupation number
of galaxies $\skaco{N_{\rm gal}}(m)/\bar{n}_{\rm gal}$, 
as discussed in \S \ref{gal}, 
we obtain the corresponding approximations for computing the 2- and 3-halo
contributions to the 3PCF of galaxies. 

\subsection{2D 3-point correlation function}

It is straightforward to extend the approximations for the 3-dimensional 3PCF 
to the 3PCF of weak lensing fields. As discussed for 
equation (\ref{eqn:f2d3pt}), combining equation (\ref{eqn:app2h}) 
with Limber's equation yields the following approximation for the 2-halo term
of the weak lensing convergence: 
\begin{eqnarray}
Z^{\rm app}_{2h}(\bm{r}_1,\bm{r}_2,\bm{r}_3)
&=&\int^{\chi_s}_0\!\!d\chi W^3(\chi,\chi_s)d_A^{-4}
\left[\int\!\!dm~ n(m)\left(\frac{m}{\bar{\rho}_0}\right)^2 b(m)
\int\!\!\frac{l_1dl_1}{2\pi} (\tilde{u}_{m}(k_1))^2J_0(l_1r_{12})
\right.\nonumber\\
&&\times \left.\int\!\!dm'~ n(m')\left(\frac{m'}{\bar{\rho}_0}\right)
b(m')\int\!\!\frac{l_2dl_2}{2\pi}
\tilde{u}_{m'}(k_2)P^L(k_2; \chi)J_0(l_2 r_{23})+\mbox{\rm cyc.}\right],
\label{eqn:app2d2h}
\end{eqnarray}
where $k_i=l_i/d_A(\chi)$. Note that the redshift dependence of 
$P^L(k; \chi)$ is given by the linear growth factor $D(z)$. 

Likewise, the approximation for the 3-halo term can be obtained by
the following replacements in equation (\ref{eqn:app3h}) and the additional 
integration over the redshift:
\begin{eqnarray}
(\Xi, \Phi\rightarrow \Xi_{2D}, \Phi_{2D}) 
\mbox{ in } \zeta^{\rm app}_{3h}, 
\label{eqn:app2d3h}
\end{eqnarray}
where $\mu_{1323}\equiv (\bm{r}_{13}\cdot \bm{r}_{23})/(r_{13}r_{23})$ and 
\begin{eqnarray}
\Xi^{(\mu)}_{2D}(r)&\equiv& \int\!\!\frac{ldl}{2\pi } \left[\int\!\!dm~ 
n(m)\left(\frac{m}{\bar{\rho}_0}\right) b(m)u_m(k)\right]^\mu 
P^L(k)J_0(lr), \nonumber \\
\Phi^{(\mu)}_{2D}(r)&\equiv &\int\!\!\frac{k^2dk}{2\pi^2 } \left[\int\!\!dm~ 
n(m)\left(\frac{m}{\bar{\rho}_0}\right) b(m)u_m(k)\right]^\mu 
\frac{P^L(k)}{k^2}J_0(lr). 
\end{eqnarray}

\section{Approximations for the 1-halo term of the 3PCF}
\label{app}

In this section, we present approximations for calculating the 1-halo
contribution to the 3D or 2D 3PCF with 
the Fourier-space halo model. These are combined with the approximation in
TJ02 for calculating the skewness parameter of the density field. 
 
\subsection{3D 3-point correlation function}

Based on the Fourier-space halo approach, the 1-halo contribution to the
3PCF of the density field can be obtained from equations
(\ref{eqn:bisp}) and (\ref{eqn:r3pt}) as
\begin{equation}
\zeta_{1h}(\bm{r}_1,\bm{r}_2,\bm{r}_3)=\int\!\!dm~ n(m)
\left(\frac{m}{\bar{\rho}_0}\right)^3 \int\!\!\prod_{i=1}^{3}
\frac{d^3\bm{k}_i}{(2\pi)^3}\tilde{u}_m(\bm{k}_i)
\exp[i\bm{k}_i\cdot\bm{r}_i](2\pi)^3\delta_D(\bm{k}_{123}).
\label{eqn:f3ptdef}
\end{equation}
The Dirac-delta function allows us to eliminate one of the three wavevectors
$\bm{k}_i$.  For example, let us eliminate
$\bm{k}_3$, leading to the replacement $\bm{k}_3=-\bm{k}_{12}$ in the
above equation.  Then, the integrand has 
dependences on $d^3\bm{k}_i$ via $\tilde{u}_m(k_{12})$ and
$\exp[i\bm{k}_i\cdot\bm{r}_i]$.  TJ02 proposed that the angular
dependence of $\tilde{u}_m(k_{12})$ can be approximated with the
replacement:
\begin{equation}
\tilde{u}_m(k_{12})\longrightarrow \tilde{u}_m(\tilde{k}_{12}),
\end{equation}
where $\tilde{k}_{12}=(k_1^2+k_2^2-k_1k_2)^{1/2}$. TJ02 showed that this
replacement works well for predictions of the skewness parameters of the
3D density field and the weak lensing field\footnote{Note that, although one
might imagine another possible replacement 
$\tilde{u}_m(k_{12})\rightarrow \tilde{u}_m(k_1)$,  
it overestimates the true value by about $\sim 50\%$ for the 1-halo term, as
shown in TJ02.}.
If we assume that this
 holds for the 3PCF calculation, we can analytically perform the angular
integrations in $\int\, d^3\bm{k}_i$, to get
\begin{equation}
\zeta_{1h}(\bm{r}_{1},\bm{r}_2,\bm{r}_3)\approx
F_{\zeta}(r_{13},r_{23})\equiv 
\int\!\!dm~ n(m)\left(\frac{m}{\bar{\rho}_0}\right)^3
\int\!\!\frac{dk_1}{2\pi^2}\frac{dk_2}{2\pi^2}~
k_1^2k_2^2 \tilde{u}_m(k_1)\tilde{u}_m(k_2)\tilde{u}_m(\tilde{k}_{12})
j_0(k_1r_{13})j_0(k_2r_{23}). 
\end{equation}
The above approximation reduces the 7-dimensional integral in
equation (\ref{eqn:f3ptdef}) to a 3-dimensional one.  
However, the resulting 3PCF
depends only on the two side lengths $r_{13}$ and $r_{23}$, 
and thus has no dependence on the interior
angles.  If we had started by eliminating $\bm{k}_2$ or $\bm{k}_1$ in equation
(\ref{eqn:f3ptdef}), the resulting 3PCF would have
dependences only on the two side lengths $r_{12}$ and $r_{23}$ or
$r_{12}$ and $r_{13}$, respectively.  Thus, the
approximate result for the 3PCF has different forms even for the
same triangle shape. To resolve this, we propose the following 
symmetrized form as the approximation for calculating $\zeta_{1h}$:
\begin{equation}
\zeta_{1h}^{\rm app}(\bm{r}_1,\bm{r}_2,\bm{r}_3)=\frac{1}{3}
\left[F_{\zeta}(r_{12},r_{13})+F_\zeta(r_{12},r_{23})
+F_\zeta(r_{13},r_{23})\right],
\label{eqn:app1h} 
\end{equation}
The dependences of this form on the three parameters of the triangle
shape, $r_{12}$, $r_{23}$ and $r_{31}$, are like the hierarchical form
$\zeta=
Q[\xi(r_{12})\xi(r_{31})+\xi(r_{12})\xi(r_{23})+\xi(r_{13})\xi(r_{22})]$.
The factor $1/3$ is simply chosen so that it agrees with 
the third-order moment given in TJ02 for the limit 
$r,q\rightarrow 0$. As shown in \S
\ref{appperform}, the approximation (\ref{eqn:app1h}) is accurate
compared with the exact value computed from equation (\ref{eqn:r3pt}),
even though each term on the r.h.s is far from the exact value (see 
Figure \ref{fig:appq}).

\subsection{2D 3-point function}

In the spirit of equation (\ref{eqn:app1h}), the corresponding 
approximation for the projected 3PCF given by equation (\ref{eqn:r2d3pt}) 
can be constructed as:
\begin{equation}
Z^{\rm app}_{1h}(\bm{\theta}_1,\bm{\theta}_2,\bm{\theta}_3)
=\frac{1}{3}\left[F_Z(\theta_{12},\theta_{13})+F_Z(\theta_{21},\theta_{23})
+F_Z(\theta_{31},\theta_{32})\right],
\label{eqn:2dapp1h}
\end{equation}
with
\begin{equation}
F_Z(\theta,\varphi)\equiv 
\int^{\chi_s}_0\!\!d\chi~ W^3(\chi) d_A^{-4}(\chi)
\int\!\!dm~ n(m)\left(\frac{m}{\bar{\rho}_0}\right)^3
\int\frac{l_1dl_1}{2\pi}\frac{l_2dl_2}{2\pi}~ 
\tilde{u}_m(k_1)\tilde{u}_m(k_2)\tilde{u}_m(\tilde{k}_{12})J_0(\theta l_1)
J_0(\varphi l_2), 
\end{equation}
where $k_i=l_i/d_A(\chi)$.  The 4-dimensional integration required
to carry out the above approximation is the same level of complexity
as the real-space halo model expression in equation (\ref{eqn:r2d3pt}).



\label{lastpage}
\end{document}